\documentclass[preprint,preprintnumbers,amsmath,amssymb]{revtex4-1}
\usepackage{amsfonts}    
\usepackage{amssymb}
\usepackage{amsmath}
\usepackage{latexsym}
\usepackage{eepic}
\usepackage{graphicx}
\usepackage[usenames,dvipsnames]{color}
\usepackage{color}

\usepackage[active]{srcltx}

\newcommand{\pa}{\partial}

\newcommand{\Om}{\Omega}
\newcommand{\om}{\omega}

\newcommand{\De}{\Delta}

\newcommand{\rar}{\rightarrow}
\newcommand{\lrar}{\leftrightarrow}

\newcommand{\non}{\nonumber}

\begin{document}

\title{Four-body problem in $d$-dimensional space: ground state, (quasi)-exact-solvability. IV}

\author{
M.A.~Escobar-Ruiz,\\[8pt]
Centre de Recherches Math\'ematiques, Universit\'e de Montr\'eal, \\
C.P. 6128, succ. Centre-Ville, Montr\'eal, QC H3C 3J7, Canada\\[8pt]
escobarr@crm.umontreal.ca\\[8pt]
     Willard Miller, Jr.\\[8pt]
School of Mathematics, University of Minnesota, \\
Minneapolis, Minnesota, U.S.A.\\[8pt]
miller@ima.umn.edu\\
[10pt]
and \\[10pt]
Alexander V Turbiner\\[8pt]
Instituto de Ciencias Nucleares, UNAM, M\'exico DF 04510, Mexico\\[8pt]
turbiner@nucleares.unam.mx
}

\begin{abstract}
Due to its great importance for applications, we  generalize and extend the approach of our previous papers to study aspects of the quantum and classical  dynamics of a $4$-body system with equal masses in {\it $d$}-dimensional space with interaction depending only on mutual (relative) distances.
The study is restricted to solutions in the space of relative motion which are functions of mutual (relative) distances only. The ground state (and some other states) in the quantum case and some trajectories in the classical case are of this type. We construct the quantum Hamiltonian for which these states are eigenstates.  For $d \geq 3$, this describes a six-dimensional quantum particle moving in a curved space with special $d$-independent metric in a certain $d$-dependent singular potential, while for $d=1$ it corresponds to a three-dimensional particle and coincides with the $A_3$ (4-body) rational Calogero model; the case $d=2$ is exceptional and is discussed separately. The kinetic energy of the system has a hidden $sl(7,{\bf R})$ Lie (Poisson) algebra structure, but  for the special case $d=1$ it becomes degenerate with hidden algebra $sl(4,R)$.  We find an exactly-solvable four-body $S_4$-permutationally invariant, generalized harmonic oscillator-type potential as well as a quasi-exactly-solvable four-body sextic polynomial type potential with singular terms. Naturally, the tetrahedron whose vertices correspond to the positions of the particles provides pure geometrical variables, volume variables, that lead to exactly solvable models. Their generalization to the $n$-body system as well as the case of non-equal masses is briefly discussed.

\end{abstract}

\maketitle

\newpage

\section*{Introduction}

Consider  four classical particles  in $d$-dimensional space with potential depending on mutual relative distances alone.  After separation of the center-of-mass motion, and assuming zero total (relative) angular momentum, the trajectories are defined by evolution of the relative (mutual) distances. It is an old question to find equations for trajectories which depend on relative distances only; in the three-body case this problem can be traced back to J-L Lagrange (1772). In general, this problem was solved for three-body case in \cite{TME3-3,TME3-d}. Naturally, the vector positions of four particles in a three-dimensional space form a tetrahedron, the corresponding edges are nothing but the six relative distances between the particles. Thus, we can formulate the problem in terms of the evolution of such a geometrical object. We call it the \emph{tetrahedron of interaction}.

The aim of the present paper is to construct the four-body Hamiltonian which depends on the six relative distances and describes the motion of the tetrahedron of interaction in $d$-dimensional space. Our strategy is to study the quantum problem first for $d \geq 3$. Then, using geometrical variables obtained from the tetrahedron, we impose constraints on the edges (relative distances) and faces to degenerate the Hamiltonian to the planar $d=2$ and one-dimensional $d=1$ cases. The corresponding classical Hamiltonian is obtained through the {\it de-quantization} procedure \cite{MTE:2018}, of replacement of the quantum momentum by the classical one with preservation of positivity of kinetic energy. In \cite{MTE:2018}, we studied the $n$-body system for $d\geq n-1$ while in the present paper we will introduce new geometrical variables which allow to analyze the case $d<n-1$.

The quantum Hamiltonian for four $d$-dimensional particles with translation-invariant potential,
which depends on relative (mutual) distances between particles only, is of the form
\begin{equation}
\label{Hgen}
   {\cal H}\ =\ -\Delta^{(4d)}\ +\  V(r_{ij})\ ,\
\end{equation}
$\Delta^{(4d)}  \equiv \sum_{i=1}^4 \frac{1}{2 m_i} \De_i^{(d)}$,\, here $\De_i^{(d)}$ is the $d$-dimensional Laplacian,
\[
     \De_i^{(d)}\ =\ \frac{\pa^2}{\pa{{\bf r}_i} \pa{{\bf r}_i}}\ ,
\]
associated with the $i$th body with coordinate vector ${\bf r}_i \equiv {\bf r}^{(d)}_i=(x_{i,1}\,, x_{i,2}\,,x_{i,3}\ldots \,,x_{i,d})$\,, and
\begin{equation}
\label{rel-coord}
r_{ij}\ = \ |{\bf r}_i - {\bf r}_j| \ ,\qquad i,j=1,2,3,4\ ,
\end{equation}
is the (relative) distance between particles $i$ and $j$, $r_{ij}=r_{ji}$\,. For simplicity, hereafter all masses in $\Delta^{(4d)}$ are assumed to be equal: $m_i=m=1$. The eigenvalue problem for ${\cal H}$ is defined on the configuration space ${\mathbb R}^{4d}$.

\begin{figure}[h]
  \centering
   \includegraphics[width=13.0cm]{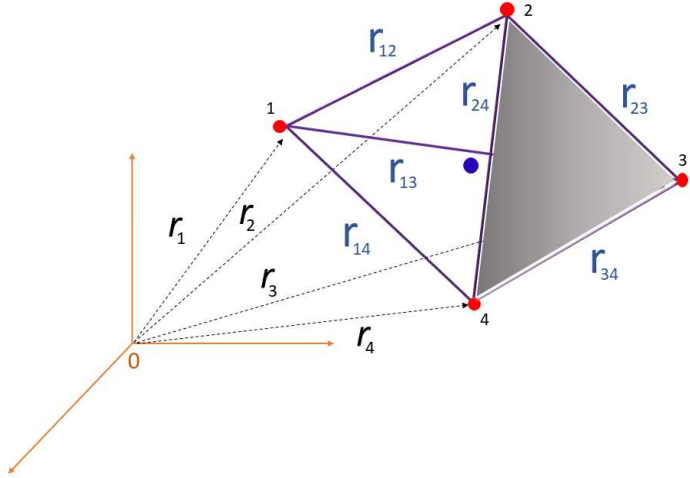}
  \caption{Four-body system: at $d=3$, the coordinate vectors ${\bf r}_i$ mark positions of vertices of the tetrahedron of interaction with sides $r_{ij}$. For illustration one of the faces of
  this tetrahedron (shaded triangle) and the center-of-mass (blue large bubble) are marked.}
\label{Fig}
\end{figure}

The number of relative distances $r_{ij}$ is equal to the number of edges of the tetrahedron
which is formed by taking the particles' positions as vertices. We call this tetrahedron
the {\it tetrahedron of interaction},
see for illustration Fig.\ref{Fig}.

The center-of-mass motion described by vectorial coordinate
\[
    {\bf R}_{CM} \ =\ \frac{1}{{\sqrt 4}}\,\sum_{k=1}^{4} {\bf r}_{_k}\ ,
\]
can be separated out; this motion is described by a $d$-dimensional plane wave, $\sim e^{i\, {\bf k}\cdot {\bf R}{_{CM}}}$.

The spectral problem is formulated in the space of relative motion ${\Re}_{rel} \equiv {\mathbb{R}}^{3d}$; it is of the form,
\begin{equation}
\label{Hrel}
   {\cal H}_{rel}\,\Psi(x)\ \equiv \ \bigg(- \frac{1}{2}\,\De_{rel}^{(3d)} + V(r_{ij})\bigg)\, \Psi(x)\ =\ E \, \Psi(x)\ ,\quad
   \Psi \in L_2 ({\Re}_{rel} )\ ,
\end{equation}
where $\De_{rel}^{(3d)}$ is the flat-space Laplacian in the space of relative motion.

If the space of relative motion ${\Re}_{rel}$ is parameterized by three, $d$-dimensional vectorial Jacobi coordinates
\begin{equation}
\label{rj}
     {\bf q}_{j} \ = \ \frac{1}{\sqrt{j(j+1)}}\sum_{k=1}^j k\,({\bf r}_{k+1} - {\bf r}_{{k}})\ ,
        \qquad\qquad j=1,2,3\ ,
\end{equation}
the flat-space $3d$-dimensional Laplacian in the space of relative motion becomes diagonal
\begin{equation}
\label{Dflat}
       \De_{rel}^{(3d)}\ =\ \sum_{j=1,2} \frac{\pa^2}{\pa{{\bf q}_j} \pa{{\bf q}_j}}\ .
\end{equation}
Thus, ${\bf q}_{j}$ plays a role of the Cartesian coordinate vector in the space of relative motion.

The cases $d=2$ (four bodies on a plane) and $d=1$ (four bodies on a line) are special. For $d=2$ the \emph{tetrahedron of interaction} degenerates either into a quadrangle with four external vertices or
triangle with three external vertices and one internal (in both cases the volume of tetrahedron vanishes, it plays a role of a constraint).
For $d=1$ it degenerates into interval: the vertices of the tetrahedron correspond to two endpoints and two marked points inside the interval, the volume of tetrahedron as well as the areas of all faces (triangles) are equal to zero identically. It implies that on the line ($d=1$) the relative variables obey three constraints
\begin{equation}
\label{rel3}
   x_{12} + x_{31} +  x_{23}  \ =\ 0\ , \qquad  x_{13} + x_{41} +  x_{34}  \ =\ 0 \ , \qquad  x_{12} + x_{24} +  x_{41}  \ =\ 0 \ ,
\end{equation}
where it is assumed that $x_i$ denotes the position of the $i$th body and $x_{ij}=x_i-x_j$.
Hence, the six relative distances are related and only three of them are independent. Therefore, see \cite{RT:1996}
\begin{equation}
\label{Drel3-1}
   \De_{rel}^{(3)}\ =\ 2\,\bigg(\frac{\pa^{2}}{\pa x_{12}^{2}}\ +\
   \frac{\pa^{2}}{\pa x_{13}^{2}}\ +\
   \frac{\pa^{2}}{\pa x_{14}^{2}}\ +\
   \frac{\pa^{2}}{\pa x_{12}\,\pa x_{13}} + \frac{\pa^{2}}{\pa x_{12} \,\pa x_{14}} + \frac{\pa^{2}}{\pa x_{13} \,\pa x_{14}}\bigg) \ .
\end{equation}


cf. (\ref{Dflat}). The configuration space ${\Re}_{rel}$ is $0 \, < \,  x_{12}  < \,  x_{13} < \,  x_{14}   \, < \, \infty$. Now,

{\large \it Observation} \cite{Ter}\,:
\begin{quote}
  There exists a family of the eigenstates of the Hamiltonian (\ref{Hgen}), including the ground state,
  which depends on six relative distances $\{r_{ij}\}$ only\,. The same is correct for the $n$ body problem:
  there exists a family of the eigenstates, including the ground state, which depends on  relative distances only.
\end{quote}

\bigskip

In a way this observation is presented for the case of scalar particles, bosons. It can be generalized to the case of fermions, namely:
\bigskip

\begin{quote}
   In the case of four fermions there exists a family of the eigenstates of the Hamiltonian (\ref{Hgen}), including the ground state, in which the coordinate functions depend on six relative distances $\{r_{ij}\}$ only\,. The same is correct for the $n$ body problem, see \cite{MTE:2018}:
   there exists a family of the eigenstates, including the ground state, in which the coordinate functions depend on relative distances only.
\end{quote}

\bigskip

\noindent
Our primary goal is to find the differential operator, in the form of the Hamiltonian with positive-definite kinetic energy, in the space of relative distances $\{r_{ij}\}$ for which these states are eigenstates. In other words, to find a differential equation depending only on $\{r_{ij}\}$ for which these states are solutions. This  implies a study of the evolution of the tetrahedron of interaction with fixed center-of-mass. We consider the case of four spinless particles.


\section{Generalities}

As a first step let us change the variables (\ref{rj}) in the space of relative motion ${\Re}_{rel}$ :
\[
({\bf q}_{j}) \ \lrar \ (\,\{r_{ij}\},\,\{ \Om\}\,) \ ,
\]

it is a generalization of the Euler coordinates; where for $d>2$ the number of (independent) relative distances $\{r_{ij}\}$ is equal to 6 and $\{\Om\}$ is a collection of $(3d-6)$ angular variables. Thus, we split ${\Re}_{rel}$ into a combination of the space of relative distances ${\Re_{radial}}$ and a space parameterized by angular variables, essentially those on the sphere $S^{3(d-2)}$. There are known several ways to introduce variables in ${\Re}_{rel}$, see e.g. \cite{Gu}. In particular, unlike \cite{Gu}, for the space of relative distances ${\Re_{radial}}$ we take the relative (mutual) distances $r_{ij}$.

A key observation is that in new coordinates $(\{r_{ij}\},\,\{ \Om\})$ the flat-space Laplace operator,
the kinetic energy operator in (\ref{Hrel}), takes the form of the sum of two second-order differential
operators
\begin{equation}
\label{addition}
    \frac{1}{2}\De_{rel}^{(3d)}\ =\ {\De_{radial}^{(6)}}(r_{ij}, \pa_{ij}) \ + \
    {\De}_{\Om}^{(3d-6)} (r_{ij}, \Om, \pa_{ij}, \pa_{\Om})
    \ ,\qquad \pa_{ij} \equiv \frac{\pa}{\pa_{r_{ij}}}\ ,
\end{equation}
($d>2$) where the first operator depends on relative distances {\it only} \footnote{Hence,  it contains derivatives w.r.t. relative distances while the coefficient functions do not depend on angles},  while the second operator depends on angular derivatives in such a way that it annihilates any angle-independent function $\Psi$, namely
\[
  {\De}_\Om^{(3d-6)} (r_{ij}, \Om, \pa_{ij}, \pa_{\Om})\, \Psi(r_{ij})\ =\ 0\ .
\]
Hereafter, we omit the superscripts in ${\De_{radial}^{(6)}}$, $\De_{rel}^{(3d)}$ and ${\De}_{\Om}^{(3d-6)}$.

The special cases $d=1$ and $d=2$ will be considered separately in section IV. In particular, for $d=1$ the operator ${\De_\Om}$ is absent (no angular variables occur), thus
\[
  \De_{radial}(r_{ij}, \pa_{ij})\ =\ \frac{1}{2}\De_{rel}(r_{ij}, \pa_{ij})\ ,
\]
see (\ref{Drel3-1}).
For $d>2$, the commutator $$[{\De_{radial}}\, ,\, {\De}_\Om ] \neq 0\ .$$

Now, if we look for angle-independent solutions of (\ref{Hrel}), due to the decomposition (\ref{addition}) the general spectral problem (\ref{Hrel}) is reduced to a particular spectral problem
\begin{equation}
\label{Hrel-Mod}
   {{\cal H}}_r\,\Psi(r_{ij})\ \equiv \ \bigg(- {\De_{radial}}(r_{ij}, \pa_{ij}) + V(r_{ij})\bigg)\, \Psi(r_{ij})\ = \ E\,\Psi(r_{ij})\ ,\qquad
   \Psi \in L_2 ({\Re_{{radial}}})\ ,
\end{equation}
where ${\Re_{radial}\subset \Re_{rel}}$ is the space of relative distances. Clearly, we can write
\begin{equation}
\label{Drgij}
     {\De_{radial}}(r_{ij}, \pa_{ij})\ =\ g^{\mu \nu}(r) \pa_{\mu} \pa_{\nu}\ +\ b^{\mu}(r) \pa_{\mu} \quad , \qquad (\mu,\, \nu=1,2,3,4,5,6) \ ,
\end{equation}
where $g^{\mu \nu}(r)$ is a $6\times 6$ matrix whose entries are basically the coefficients in front of the second derivatives $\pa_{\mu} \pa_{\nu}$, and $b^{\mu}(r)$ is a column vector; both are $r$-dependable. In (\ref{Drgij}), we made the identifications $1\rightarrow r_{12}$,\, $2\rightarrow r_{13}$,\, $3\rightarrow r_{14}$,\, $4\rightarrow r_{23}$,\, $5\rightarrow r_{24}$,\, $6\rightarrow r_{34}$ for $\mu$ and $\nu$.

For any $d>2$ one can find the $d$-dependent gauge factor $\Gamma=\Gamma(r_{ij})$ such that ${\De_{radial}}(r_{ij}, \pa_{ij})$ takes the form of the Schr\"odinger operator,
\begin{equation}
\label{DLB}
     \Gamma^{-1}\,{\De_{radial}}\,(r_{ij}, \pa_{ij})\, \Gamma\ =\ {\De_{LB}}(r_{ij}) - {V_{eff}}(r_{ij}) \ \equiv \ \Delta^{}_{r,\Gamma} \ ,
\end{equation}
where $\De_{LB}(r)$ is the six-dimensional Laplace-Beltrami operator with contravariant, $d$-independent metric $g^{\mu \nu}(r)$, in general, on some non-flat, (non-constant curvature) manifold. It makes sense of the kinetic energy. Here $V_{eff}(r)$ is the $d$-dependent \emph{effective potential}. The potential ${V}_{eff}(r)$ becomes singular at the boundary of the configuration space, where the determinant $D(r)=\det g^{\mu \nu}(r)$ vanishes. It can be checked that the operator $\Delta^{}_{r}$ is Hermitian with measure $D(r)^{-\frac{1}{2}}$. Eventually, we arrive at the spectral problem for the Hamiltonian
\begin{equation}
\label{Hrel-final}
   {H}_{LB}(r)\ = \ -{\De_{LB}}(r) \ + \   V_{eff}(r)  \ + \ {V}(r)     \ ,
\end{equation}
at $d>2$ with a $d$-independent Laplace-Beltrami operator ${\De_{LB}}(r)$. It is easy to see that at $d=2$, as the consequence of the vanishing volume of the tetrahedron of interaction, the operator ${\De_{LB}}(r)$ becomes degenerate: $D(r)=\det g^{\mu \nu}(r)=0$\,. The configuration space $D \geq 0$ (equivalently, the space of relative coordinates) at $d>2$ shrinks to its boundary $D = 0$ for $d=2$.

The connection between the kinetic energy ($\Delta^{(4d)}$) in the original Hamiltonian (\ref{Hgen}) and that of the Hamiltonian (\ref{Hrel-final}) can be summarized as follows,
\[
\Delta^{(4d)}\quad {}_{\overrightarrow{\text {removal of} \ {\bf R}_{CM}} } \quad \Delta^{(3d)}_{rel} \quad {}_{\overrightarrow{\text { angle-independent solutions} }} \quad \Delta^{}_{radial} \quad {}_{\overrightarrow{\text {gauge transformation}\ \Gamma}} \quad \Delta^{}_{LB} \ .
\]
Consequently, we reduce the original $4d$-dimensional problem to a six dimensional one. As for the potential, we simply add to the original $V$ the effective potential $V_{eff}$ arising from the $d-$dependent gauge transformation $\Gamma$.
Again, the case $d=1$ is special, the gauge factor is trivial, $\Gamma=1$
and
\begin{equation}
\label{Hrel1d}
     {\De_{LB}}(r) \ =\ {\De_{radial}}(r)\ =\ \De_{rel}(r)\ .
\end{equation}

Following the {\it de-quantization} procedure [1]-\cite{MTE:2018} of replacement of the quantum momentum (derivative) by the classical momentum
\[
      -i\,\pa\ \rar\ P\ ,
\]
one can get a classical analogue of the Hamiltonian (\ref{Hrel-final}),
\begin{equation}
\label{Hrel-Cl-final}
   {H}^{(c)}_{LB}(r)\ =\ g^{\mu \nu}(r)\, P_\mu\, P_\nu \ + \  V(r) \  + \  V_{eff}(r)\ .
\end{equation}
It describes the internal motion of a 6-dimensional body with tensor of inertia $(g^{\mu \nu})^{-1}$\, with center of mass  fixed.

The Hamiltonians (\ref{Hrel-final}), (\ref{Hrel-Cl-final}) are the main objects of study of this paper.


\section{Case $d=1$: concrete results }

For the one dimensional case $d=1$, we introduce the $S_4$ invariant symmetric polynomials

\begin{equation}
\label{}
\begin{aligned}
 & \sigma_1(x) \ = \ x_1 \ + \ x_2 \ + \  x_3 \  + \  x_4
\\ &  \sigma_2(x) \ = \ x_1 \,x_2 \ + \ x_1 \,x_3\ + \ x_1\, x_4\ +\ x_2\, x_3 \ + \ x_2\, x_4\ + \ x_3\, x_4
\\ &  \sigma_3(x) \ = \  x_1 \,x_2\, x_3 \ + \ x_1\, x_3\, x_4 \ +  \ x_2\, x_3\, x_4 \  +  \ x_1\, x_2\, x_4
\\ &  \sigma_4(x) \ = \   x_1\,x_2\,x_3\,x_4 \ .
\end{aligned}
\end{equation}

where it is assumed that $x_i$ denotes the position of the $i$th body.

In the variables
\[
Y \ = \ \sigma_1(x) \ , \qquad \ \tau_k \ = \  \sigma_k(y(x))\ , \qquad (k=2,3,4) \  ,
\]
here
\[
y_i(x) \ = \ x_i - \frac{1}{4}Y \ , \ \qquad (i=1,2,3,4) \ \ ,
\]
are translational invariant, the original Laplacian (\ref{Hgen}) takes the algebraic form
\begin{equation}
\label{Hgensigma}
\begin{aligned}
-\sum_{i=1}^4 \frac{1}{2} \De_i^{(d=1)}  \ & = \  -2\,\partial_{Y,Y}^2 \ + \  \tau_2\,\partial_{2,2}^2 \ + \  (2\,\tau_4-\frac{1}{2}\tau_2^2)\,\partial_{3,3}^2
 \ + \  (\tau_2\,\tau_4 - \frac{3}{8}\tau_3^2)\,\partial_{4,4}^2
 \\ & \quad + \ 3\,\tau_3\,\partial_{2,3}^2 \ + \  4\,\tau_4\,\partial_{2,4}^2 \  - \  \frac{1}{2}\tau_2\,\tau_3\,\partial_{3,4}^2 \  +  \ \frac{3}{2}\partial_2 \  + \  \frac{1}{4}\,\tau_2\,\partial_{4}  \ ,
\end{aligned}
\end{equation}
which upon the extraction of the center-of-mass motion can be rewritten in terms of the generators of algebra $sl(4,\,\bf{R})$. Moreover, it can be immediately seen that such operator describes the kinetic energy of relative motion of the four-body ($A_3$) rational Calogero model with potential
\begin{equation}
\label{VA3}
V_{A_3} \ =\ g\,\bigg( \frac{1}{x^2_{12}} \, + \, \frac{1}{x^2_{13}} \, + \,\frac{1}{x^2_{14}} \, + \,\frac{1}{x^2_{23}} \, + \,\frac{1}{x^2_{24}} \, + \,\frac{1}{x^2_{34}} \bigg) \ ,
\end{equation}
in algebraic form, $g$ is the coupling constant and $x_{ij}\equiv x_i-x_j$.

Also, for $d=1$ there exists another polynomial change of variables. In the space of relative distances the Laplace-Beltrami operator (\ref{Hrel1d}) is given by
\[
  {\De_{LB}}\ =\ \frac{\pa^{2}}{\pa x_{12}^{2}}\ +\
   \frac{\pa^{2}}{\pa x_{13}^{2}}\ +\
   \frac{\pa^{2}}{\pa x_{14}^{2}}\ +\
   \frac{\pa^{2}}{\pa x_{12}\,\pa x_{13}}\  + \ \frac{\pa^{2}}{\pa x_{12} \,\pa x_{14}} \ + \  \frac{\pa^{2}}{\pa x_{13} \,\pa x_{14}} \ ,
\]
see (\ref{Drel3-1}). It  corresponds to the three-dimensional flat space Laplacian and is evidently an algebraic operator.
Formally, it is not $S_4$ invariant unlike the original $4d$-Laplacian in (\ref{Hgen}) with $d=1$. However, the kinetic energy do
remain $S_3$ invariant.
As a realization of this $S_3$ invariance in ${\De_{LB}}$ (\ref{Hrel1d}) let us introduce the natural variables
\begin{equation}
\label{d1-xi}
 \xi_1 \ = \  x_{12} + x_{13} + x_{14} \quad ,\qquad \xi_2 \ = \   x_{12} \, x_{13} + x_{12} \, x_{14}  +  x_{13} \, x_{14}  \quad ,\qquad \xi_3 \ = \   x_{12} \, x_{13} \, x_{14}   \ ,
\end{equation}
which is a polynomial change of variables, thus (\ref{Hrel1d}) becomes
\begin{equation}
\label{Drel3-1-xi}
\begin{aligned}
  {\De_{LB}}(\xi)\ = & \   6\,\pa^2_{\xi_1} \ +\ (3\,\xi_1^2-\xi_2)\,\pa^2_{\xi_2}\ + \ (\xi_2^2-\xi_1\,\xi_3)\,\pa^2_{\xi_3}
\  +  \  8\,\xi_1\,\pa^2_{\xi_1,\xi_2 }
\\ &
  +  \ 4\,\xi_2\,\pa^2_{\xi_1,\xi_3 } \ + \ 3(\xi_1\,\xi_2-\xi_3 )\,\pa^2_{\xi_2,\xi_3 } \ + \ 3\,\pa_{\xi_2} \ + \ \xi_1\,\pa_{\xi_3}      \ .
\end{aligned}
\end{equation}
The operator (\ref{Drel3-1-xi}) is algebraic, it can be rewritten in terms of the
generators of the maximal affine subalgebra $b_3$ of the algebra $sl(4,{\bf R})$ in $\xi$-variables, c.f. below (\ref{sl4R}), see  \cite{RT:1996,Turbiner:1998}.
\begin{eqnarray}
\label{sl3R}
 {\cal J}_i^- &=& \frac{\pa}{\pa \xi_i}\ ,\qquad \quad i=1,2,3\ , \non  \\
 {{\cal J}_{ij}}^0 &=&  \xi_i \frac{\pa}{\pa \xi_j}\ , \qquad i,j=1,2,3 \ , \\
 {\cal J}^0(N) &=& \sum_{i=1}^{3} \xi_i \frac{\pa}{\pa \xi_i}-N\, , \non \\
 {\cal J}_i^+(N) &=& \xi_i\, {\cal J}^0(N)\ =\
    \xi_i\, \left( \sum_{j=1}^{3} \xi_j\frac{\pa}{\pa \xi_j}-N \right)\ ,
       \quad i=1,2,3\ ,
\end{eqnarray}
where $N$ is a parameter.

\section{Case $d > 2$: concrete results }

\subsection{$r$-representation}

Assuming $d > 2$, after straightforward calculations the operator ${\De_{radial}}(r_{ij}, \pa_{ij})$ (in decomposition (\ref{addition})) can be found to be
\begin{equation}
\begin{aligned}
\label{addition3-3r}
& 2\,{\De_{radial}}(r_{ij}, \pa_{ij})\ =   \ \bigg[\ 2\,(\pa^{2}_{r_{12}} \, + \, \pa^{2}_{r_{13}} \, + \, \pa^{2}_{r_{14}} \, + \, \pa^{2}_{r_{23}}\, + \, \pa^{2}_{r_{24}}\, +  \,   \pa^{2}_{r_{34}}) + \frac{2(d-1)}{r_{12}}\,\pa_{r_{12}}
\\ &
\ + \ \frac{2(d-1)}{r_{13}}\,\pa_{r_{13}} \, +  \, \frac{2(d-1)}{r_{14}}\,\pa_{r_{14}} \, + \, \frac{2(d-1)}{r_{23}}\,\pa_{r_{23}} \, + \,  \frac{2(d-1)}{r_{24}}\,\pa_{r_{24}}\, + \, \frac{2(d-1)}{r_{34}}\,\pa_{r_{34}}
\\ &
\  + \ \frac{r_{12}^2+r_{13}^2-r_{23}^2}{r_{12}\, r_{13}}\,\pa_{r_{12}}\pa_{r_{13}} \ + \ \frac{r_{12}^2+r_{14}^2-r_{24}^2}{r_{12}\, r_{14}}\,\pa_{r_{12}}\pa_{r_{14}} \  + \ \frac{r_{13}^2+r_{14}^2-r_{34}^2}{r_{13}\, r_{14}}\,\pa_{r_{13}}\pa_{r_{14}}
\\  &
\ + \ \frac{r_{12}^2+r_{23}^2-r_{13}^2}{r_{12}\, r_{23}}\,\pa_{r_{12}}\pa_{r_{23}}\ + \ \frac{r_{12}^2+r_{24}^2-r_{14}^2}{r_{12}\, r_{24}}\,\pa_{r_{12}}\pa_{r_{24}}\ + \ \frac{r_{23}^2+r_{24}^2-r_{34}^2}{r_{23}\, r_{24}}\,\pa_{r_{23}}\pa_{r_{24}}
\\  &
\ + \ \frac{r_{13}^2+r_{23}^2-r_{12}^2}{r_{13}\, r_{23}}\,\pa_{r_{13}}\pa_{r_{23}}\  + \ \frac{r_{13}^2+r_{34}^2-r_{14}^2}{r_{13}\, r_{34}}\,\pa_{r_{13}}\pa_{r_{34}} \ + \ \frac{r_{23}^2+r_{34}^2-r_{24}^2}{r_{23}\, r_{34}}\,\pa_{r_{23}}\pa_{r_{34}}
\\  &
\ + \ \frac{r_{14}^2+r_{24}^2-r_{12}^2}{r_{14}\, r_{24}}\,\pa_{r_{14}}\pa_{r_{24}} \ + \ \frac{r_{14}^2+r_{34}^2-r_{13}^2}{r_{14}\, r_{34}}\,\pa_{r_{14}}\pa_{r_{34}} \ + \ \frac{r_{24}^2+r_{34}^2-r_{23}^2}{r_{24}\, r_{34}}\,\pa_{r_{24}}\pa_{r_{34}} \ \bigg] \ .
\end{aligned}
\end{equation}
Notice the absence of the cross terms $\pa_{r_{12}}\pa_{r_{34}}$, $\pa_{r_{13}}\pa_{r_{24}}$ and $\pa_{r_{14}}\pa_{r_{23}}$, each of them involves two disconnected edges of the tetrahedron of interaction.

In general, the operator (\ref{addition3-3r}) does not depend on the choice of the angular variables $\Om$, but the operator ${\De}_\Om (r_{ij}, \pa_{ij}, \Om, \pa_{\Om})$ in (\ref{addition}) does.
The configuration space in the space of relative distances is
\begin{equation}
\label{CFr}
 0 < \  r_{a} < \ r_{b} + r_{c} < \ \infty  ,\qquad 0 < \ r_{b}< r_{a}+r_{c}< \ \infty,\qquad 0 < \ r_{c}< r_{a}+r_{b}< \ \infty\ ,
\end{equation}
($ a \neq b \neq c = 12,13,14,23,24,34 $).

\subsection{$\rho$-representation}

Formally, the operator (\ref{addition3-3r}) is invariant under reflections $Z_2 \oplus Z_2 \oplus Z_2\oplus Z_2 \oplus Z_2\oplus Z_2 $,
\[
r_{12} \Leftrightarrow -r_{12}  \ ,\quad  r_{13} \Leftrightarrow -r_{13} \ ,\quad  r_{14} \Leftrightarrow -r_{14} \ ,\quad r_{23} \Leftrightarrow -r_{23}  \ ,\quad  r_{24} \Leftrightarrow -r_{24} \ ,\quad  r_{34} \Leftrightarrow -r_{34} \ .
\]
If we introduce new reflection invariant variables,
\begin{equation}
\label{rho}
r_{12}^2\ =\ \rho_{12}\ ,\quad r_{13}^2\ =\ \rho_{13}\ ,\quad r_{14}^2\ =\ \rho_{14} \ , \quad r_{23}^2\ =\ \rho_{23}\ ,\quad r_{24}^2\ =\ \rho_{24}\ ,\quad r_{34}^2\ =\ \rho_{34}\ ,
\end{equation}
the operator (\ref{addition3-3r}) becomes algebraic,

\[
{\De_{radial}}(\rho_{ij}, \pa_{ij})\ =  \   4(\rho_{12} \,\pa^2_{\rho_{12}} + \rho_{13}\, \pa^2_{\rho_{13}} +\rho_{14}\, \pa^2_{\rho_{14}}+\rho_{23}\, \pa^2_{\rho_{23}} +\rho_{24}\, \pa^2_{\rho_{24}} +\rho_{34}\, \pa^2_{\rho_{34}} )
\]
\[
+  2 \bigg((\rho_{12} + \rho_{13} - \rho_{23})\pa_{\rho_{12}}\pa_{\rho_{13}}\ +
          (\rho_{12} + \rho_{14} - \rho_{24})\pa_{\rho_{12}}\pa_{\rho_{14}}\ +
          (\rho_{13} + \rho_{14} - \rho_{34})\pa_{\rho_{13}}\pa_{\rho_{14}} \bigg)
\]
\[
+  2 \bigg((\rho_{12} + \rho_{23} - \rho_{13})\pa_{\rho_{12}}\pa_{\rho_{23}}\ +
          (\rho_{12} + \rho_{24} - \rho_{14})\pa_{\rho_{12}}\pa_{\rho_{24}}\ +
          (\rho_{23} + \rho_{24} - \rho_{34})\pa_{\rho_{23}}\pa_{\rho_{24}}
    \bigg)
\]
\[
+  2 \bigg((\rho_{13} + \rho_{23} - \rho_{12})\pa_{\rho_{13}}\pa_{\rho_{23}}\ +
          (\rho_{13} + \rho_{34} - \rho_{14})\pa_{\rho_{13}}\pa_{\rho_{34}}\ +
          (\rho_{23} + \rho_{34} - \rho_{24})\pa_{\rho_{23}}\pa_{\rho_{34}}
    \bigg)
\]
\[
+  2 \bigg((\rho_{14} + \rho_{24} - \rho_{12})\pa_{\rho_{14}}\pa_{\rho_{24}}\ +
          (\rho_{14} + \rho_{34} - \rho_{13})\pa_{\rho_{14}}\pa_{\rho_{34}}\ +
          (\rho_{24} + \rho_{34} - \rho_{23})\pa_{\rho_{24}}\pa_{\rho_{34}}
    \bigg)
\]
\begin{equation}
\label{addition3-3rho}
 +\ 2\,d\,(\pa_{\rho_{12}} + \pa_{\rho_{13}}+ \pa_{\rho_{14}}+ \pa_{\rho_{23}}+ \pa_{\rho_{24}}+ \pa_{\rho_{34}}) \ .
\end{equation}
As a function of the $\rho$-variables, the operator (\ref{addition3-3rho}) is not $S_6$ permutationally-invariant. Nevertheless, it remains $S_4$ invariant under the permutations of the particles (vertices of tetrahedron of interaction). For the three-body case, where the number of $\rho$ variables (relative distances) equals the number of particles, the corresponding operator $\Delta_{radial}$ is indeed $S_3$ permutationally-invariant.

From (\ref{CFr}) and (\ref{rho}) it follows that the corresponding configuration space in $\rho$ variables is given by the conditions
\[
0 < \rho_{A},\rho_{B},\rho_{C} < \infty\ ,\quad
{\rho}_{A} <  (\sqrt{{\rho}_{B}} + \sqrt{{\rho}_{C}})^2,\quad {\rho}_{B} < (\sqrt{{\rho}_{A}} + \sqrt{{\rho}_{C}})^2,\quad {\rho}_{C} < \ (\sqrt{{\rho}_{A}} + \sqrt{{\rho}_{B}})^2 \ ,
\]
($ A \neq B \neq C = 12,13,14,23,24,34 $). We remark that
\begin{equation}
\label{CFrho}
\quad
S^2_{\Delta ABC} \equiv  \frac{ 2 \,(\rho _{A}\, \rho _{B}+  \rho _{A}\, \rho _{C}+
       \rho _{B}\, \rho _{C})  - (\rho _{A}^2+\rho _{B}^2+\rho _{C}^2) }{16} \ \geq \ 0   \ ,
\end{equation}
because the left-hand side (l.h.s.) is equal to
$$\frac{1}{16}(r_{A}+r_{B}-r_{C})(r_{A}+r_{C}-r_{B})(r_{B}+r_{C}-r_{A})(r_{A}+r_{B}+r_{C})\ ,$$
and conditions (\ref{CFr}) should hold. Therefore, following the Heron formula, $S^2_{\Delta ABC}$ is the square of the area of the \emph{triangle of interaction} with sides $r_{A},\,r_B$ and $r_C$ \,. The triangles of interaction are nothing but the faces of the tetrahedron.

The associated contravariant metric for the operator ${\De_{radial}}(\rho)$ defined by coefficients in front of second derivatives is remarkably simple
{\scriptsize
\begin{equation}
\label{gmn33-rho}
 g^{\mu \nu}(\rho)\ = \left(
\begin{array}{cccccc}
 4\, \rho _{12} & \rho _{12}+\rho _{13}-\rho _{23} & \rho _{12}+\rho _{14}-\rho _{24} & \rho _{12}-\rho _{13}+\rho _{23} & \rho _{12}-\rho _{14}+\rho _{24} & 0 \\
 \rho _{12}+\rho _{13}-\rho _{23} & 4\, \rho _{13} & \rho _{13}+\rho _{14}-\rho _{34} & \rho _{13}+\rho _{23}-\rho _{12} & 0 & \rho _{13}-\rho _{14}+\rho _{34} \\
 \rho _{12}+\rho _{14}-\rho _{24} & \rho _{13}+\rho _{14}-\rho _{34} & 4 \,\rho _{14} & 0 & \rho _{14}+\rho _{24}-\rho _{12} & \rho _{14}+\rho _{34}-\rho _{13} \\
 \rho _{12}-\rho _{13}+\rho _{23} & \rho _{13}+\rho _{23}-\rho _{12} & 0 & 4 \,\rho _{23} & \rho _{23}+\rho _{24}-\rho _{34} & \rho _{23}-\rho _{24}+\rho _{34} \\
 \rho _{12}-\rho _{14}+\rho _{24} & 0 & \rho _{14}+\rho _{24}-\rho _{12} & \rho _{23}+\rho _{24}-\rho _{34} & 4\, \rho _{24} & \rho _{24}+\rho _{34}-\rho _{23} \\
 0 & \rho _{13}-\rho _{14}+\rho _{34} & \rho _{14}+\rho _{34}-\rho _{13} & \rho _{23}-\rho _{24}+\rho _{34} & \rho _{24}+\rho _{34}-\rho _{23} & 4\, \rho _{34} \\
\end{array}
\right) \ ,
\end{equation}
}
it is linear in $\rho$-coordinates(!) with positive definite factorized determinant

\begin{equation}
\label{determinant4}
D(\rho)\ =\ 36864\, F_1\,F_2\ ,
\end{equation}
where
\begin{equation}
\label{F1}
F_1\ =\ {V}_4^2\ ,
\end{equation}
\begin{equation}
\label{F2}
F_2\ =\ 36\,\tilde {V}_1^2\,{V}_4^2\ -\ \tilde {V}_2^ 2 \, \tilde {V}_3^2\ ,
\end{equation}
here
\begin{itemize}
  \item ${V}^2_4$ is the square of the volume of the tetrahedron of interaction.
  \item $\tilde {V}_3^2$ is the sum of the four areas (squared) of the faces (triangles) of the tetrahedron.
  \item $\tilde {V}_2^ 2$ is the sum of the six edges (squared) of the tetrahedron.
  \item By definition $\tilde {V}_1^2 \equiv 1$.
\end{itemize}
see (\ref{V4})-(\ref{VarP}). Hence, $F_{1,2}$ are of geometrical nature. They define the boundary of configuration space,
$F_1=0,\,F_2=0$, where the determinant (\ref{determinant4}) degenerates, vanishes.

Following the Conjecture 3 in \cite{MTE:2018}, the operator $\Delta_{radial}(\rho)$ is self-adjoint with respect to the normalized radial measure $dv_{r}$ of the form

\[
dv_{r} \ = \ V_4^{d-6}\,d\rho_{12}\,d\rho_{13}\,d\rho_{14}\,d\rho_{23}\,d\rho_{24}\,d\rho_{34}\ .
\]

\subsection{Integral}

The reduced radial Laplacian (\ref{addition3-3rho}) admits a 3-dimensional symmetry algebra with elements of the type
\[
  L(a,b,c)\ =
\]
\[
   \left (\rho_{13}\,a+\rho_{14}\,b-\rho_{23}\,a-\rho_{24}\,b\right)\pa_{\rho_{12}}+\left(\rho_{14}(\frac32\, a+\frac72\,b+3c)-\rho_{12}
   (\frac32\, a+\frac32\,b+c)+\right.
\]
\[
   \left. \rho_{23}
   (\frac32\, a+\frac32\, b+c)-\rho_{34}(\frac32\,a+\frac72\,b+3c)\right)\pa_{\rho_{13}}+\left(\rho_{12}(\frac12\, a-\frac12\, b-c)\right.
\]
\[
   +\rho_{13} (-\frac12\, a +\frac12\, b+c)+ \left.
   \rho_{24}(\frac52\, b+\frac32\, a+3c)-\rho_{34}(\frac52\, b+\frac32\, a+3c)\right)\pa_{\rho_{23}}
\]
\[
   +\bigg(c \rho_{12}-\rho_{13}(a+3b+3c)-\rho_{24}\,c +
   \rho_{34}(a+3b+3c)\bigg)
     \pa_{\rho_{14}}+\left(\rho_{12}(2b+c+a)-\right.
\]
\[
  \left.\rho_{14}(2b+c+a)-\rho_{23}(2a+3b+3c)+\rho_{34}(2a+3b+3c)\right)\pa_{\rho_{24}}+
\]
\[
  \left(\rho_{13}(\frac12\, a
      +\frac52\, b+2c)-\rho_{14}(\frac12\, a +\frac52\, b+2c)+
    \rho_{23}(\frac32 \, a+\frac32\, b+2c)\right.
\]
\begin{equation}
\label{Lparam}
\left. -\rho_{24}(\frac32\, a+\frac32\, b+2c)\right)\pa_{ \rho_{34}}\ ,
\end{equation}
where $a,b,c$ are parameters, namely, the operator $L(a,b,c)$ commutes with
$\Delta_{radial}(\rho)$.
Out of (\ref{Lparam}) let us form the three operators $\{J_1,J_2,J_3\}$,
\[
  J_1\ \equiv \ L\,\left(\frac{2\sqrt{35}}{35}, 0, -\frac{3\sqrt{35}}{35}\right)\ ,\qquad
  J_2\ \equiv \ L\,\left(-\frac{17\sqrt{210}}{420}, \frac{35\sqrt{210}}{420}, -\frac{27\sqrt{210}}{420}\right)\ ,
\]
\[
 J_3\ \equiv \ L\,\left( \frac{5\sqrt{6}}{12}, \frac{\sqrt{6}}{12}, -\frac{\sqrt{6}}{4}\right)\ .
\]
It can be checked that they satisfy the $so(3,{\bf R})$ commutation relations
\[
[J_1,J_2]\,=\,J_3\ ,\qquad [J_2,J_3]\,=\,J_1 \ , \qquad [J_3,J_1]\,=\,J_2\ .
\]
So, the symmetry algebra of $\Delta_{radial}(\rho)$ is isomorphic to $so(3,{\bf R})$.

As for the original four-body problem (\ref{Hgen}) this integral is a {\it particular} integral: it commutes with the Hamiltonian (\ref{Hgen}) over the space of relative distances ${\Re_{radial}}$ only
\[
   [{\cal H} , L]\, {\Re_{radial}}\ \rar \ 0\ .
\]
In general,  ${\cal H}$ and $L$ do not commute.

The space of 2nd order symmetries of $\Delta_{radial}(\rho)$ is much more complicated. Due to space limitations we merely summarize our results.
The space of 2nd order symmetries is the direct sum of the 6-dimensional space $D_1$  of  symmetries whose 2nd order terms  are homogeneous of order 1 in the
$\rho$ variables ($\dim D_1=6$) and the 21-dimensional space $D_2$ of symmetries whose 2nd order terms  are homogeneous of order 2 in the $\rho$ variables ($\dim D_2=21$).  Under the adjoint action of the $so(3,{\bf R})$ 1st order symmetries, $D_1$  splits into the sum of 2 irreducible subspaces: one of dimension 1 (the Hamiltonian) and one  of dimension 5. To give a brief description of these elements it is convenient to use the complex basis $\{J^0,J^+,J^-\}$ typical for $sl(2,C)$,
\[
J^0=iJ_3\,,\quad  J^+=-J_2+iJ_1\,,\quad J^-=J_2+iJ_1\ .
\]
The finite dimensional irreducible representations of $so(3,{\bf R})$ are indexed by a non-negative integer $\ell$. The corresponding irreducible subspaces have a basis of $(2\ell+1)$ elements
\[
\{f^{(\ell)}_m :\ m=\ell,\ell-1,\cdots,-\ell\}\ ,
\]
such that the action of  $so(3,{\bf R})$ is given by
\begin{equation}
\label{o(3)reps}
 J^0 f^{(\ell)}_m\ =\ m f^{(\ell)}_m\quad ,\quad
 J^\pm f^{(\ell)}_m\ =\ \left[(\ell \pm m+1)(\ell\mp m)\right]^{1/2}f^{(\ell)}_{m\pm 1}\ .
\end{equation}
For $D_1$, the basis can be computed from $f^{(0)}_0=\Delta_{radial}(\rho)$, thus taking $\ell=0$,
and the 5 basis elements can be computed from

\begin{equation}
\begin{aligned}
 &f^{(2)}_2 \ = \
-2\rho_{13}\pa^2_{\rho_{13},\rho_{13}}+(\rho_{13}+\rho_{34}-\rho_{14})\pa^2_{\rho_{34},\rho_{13}}-
\frac{1}{33}(63+46i\sqrt{6})\rho_{23} \pa^2_{\rho_{23},\rho_{23}}
\\ &
-\frac16(3-2i\sqrt{6})d\,\pa_{\rho_{12}} + \frac{1}{11}(5+4i\sqrt{6})(\rho_{13}+\rho_{12}-\rho_{23})\pa^2_{\rho_{13},\rho_{12}}-
 \frac{1}{11}(13+6i\sqrt{6})\rho_{14}
 \pa^2_{\rho_{14},\rho_{14}}
 \\ &
+ \frac{1}{11}(5+4i\sqrt{6})(\rho_{13}-\rho_{14}-\rho_{23}+\rho_{24})\pa^2_{\rho_{34},\rho_{12}}+
 \frac{1}{11}(13+6i\sqrt{6})
 (\rho_{12}-\rho_{14}-\rho_{24})\pa^2_{\rho_{24},\rho_{14}}
 \\ &
+ (\rho_{12}-\rho_{13}-\rho_{24}+\rho_{34})\pa^2_{\rho_{23},\rho_{14}}-
 (\rho_{13}+\rho_{14}-\rho_{34})\pa^2_{\rho_{14},\rho_{13}} - \frac{1}{66}(63+46i\sqrt{6})d\,\pa_{\rho_{23}}
 \\ &
+ \frac{1}{3}(4+i\sqrt{6})(\rho_{12}-\rho_{14}-\rho_{23}+\rho_{34})\pa^2_{\rho_{24},\rho_{13}}
 +\frac13(3-2i\sqrt{6})\rho_{12}\pa^2_{\rho_{12},\rho_{12}}- d\,\pa_{\rho_{13}}
 \\  &
-\frac{1}{11}(3-2i\sqrt{6})(\rho_{23}-\rho_{34}-\rho_{24})\pa^2_{\rho_{34},\rho_{24}}+
 \frac{1}{11}(27+4i\sqrt{6})(\rho_{12}-\rho_{13}-\rho_{23})\pa^2_{\rho_{23},\rho_{13}}
 \\ &
-\frac{1}{33}(15+34i\sqrt{6})(\rho_{23}+\rho_{24}-\rho_{34})\pa^2_{\rho_{24},\rho_{23}} -
 \frac{1}{11}(13+6i\sqrt{6})\rho_{34}\pa^2_{\rho_{34},\rho_{34}}
 \\ &
 + \frac{2}{11}(1+3i\sqrt{6})(\rho_{13}-\rho_{14}-\rho_{34})\pa^2_{\rho_{34},\rho_{14}} +
 \frac13(3-2i\sqrt{6}) (\rho_{12}-\rho_{14}+\rho_{24})\pa^2_{\rho_{24},\rho_{12}}
 \\ &
 - \frac{1}{33}(3+20i\sqrt{6})\,d\,\pa_{\rho_{24}}-\frac{4}{11}(4+i\sqrt{6})(\rho_{23}+\rho_{34}-\rho_{24})
\pa^2_{\rho_{34},\rho_{23}} - \frac{1}{22}(13+6i\sqrt{6})\,d\,\pa_{\rho_{34}}
 \\  &
 - \frac{2}{33}(3+20i\sqrt{6})\rho_{24}\pa^2_{\rho_{24},\rho_{24}}+
 \frac{2}{33}(9-17i\sqrt{6})(\rho_{12}-\rho_{13}+\rho_{23}) \pa^2_{\rho_{23},\rho_{12}}
 \\ &   +\frac{1}{22}(13+6i\sqrt{6})d\,\pa_{\rho_{14}}\ ,
\end{aligned}
\end{equation}
for $\ell=2$, by using equations (\ref{o(3)reps}).

We can show that these 6 basis elements for $D_1$ are pairwise commutative
and algebraically independent. Thus  the free Hamiltonian system is integrable.
However, the 6 basis symmetries fail to satisfy the algebraic conditions for a separable 
coordinate system \cite{KKM}.
Briefly, if the  coefficients of a  2nd order  symmetry operator in the $\rho$ coordinates are given by
$R^{\mu\nu}$ the eigenforms $\omega$ and eigenalues $\lambda_j$ are the solutions of the equation
\[
\sum_{\nu=1}^6 (R^{\mu\nu}-\lambda g^{\mu\nu})\omega_\nu =0,\quad  \mu=1,\cdots,6\ ,
\]
where the coefficients of the Hamiltonian are given by (\ref{gmn33-rho}). For separability the 6 basis symmetries should pairwise commute, each should admit 6 eigenvalues and the symmetries should share the {\it same} 6 eigenforms.
By a long computation one can show that the 6 basis symmetries do not have a common basis of eigenforms.

Under the adjoint action of the 1st order symmetries, $D_2$ splits into 5 irreducible subspaces, two of dimension 1  $(\ell=0)$, two of dimension 5  $(\ell=2)$ and one of dimension 9 $(\ell=4)$. The expressions for the basis symmetries are very lengthy and we do not list them here.
One of the dimension 1 subspaces has basis ${\cal J}_1=\{J_1^2+J_2^2+J_3^2\} $
and one of the dimension 5 subspaces has basis
\[
{\cal J}_2= \{J_1^2+J_2^2-2J_3^2,\ J_1^1-2J_2^2+J_3^2,\quad J_kJ_\ell+J_\ell J_k,\ 1\le \ell<k\le 3\}\ .
\]
The basis symmetry for the other 1-dimensional irreducible subspace commutes with ${\cal J}_1$ and ${\cal J}_2$.  It appears that there are no more commutative sextuplets in this full 27-dimensional space, though we do not yet have a convincing proof.
Thus, it appears that the free system is integrable, but not separable.

\subsection{The Representations of $sl(7,{\bf R})$}

In the $\rho-$representation, the operator (\ref{addition3-3rho}) is $sl(7,{\bf R})$-Lie algebraic - it can be rewritten in terms of the
generators of the maximal affine subalgebra $b_7$ of the algebra $sl(7,{\bf R})$, see e.g. \cite{Turbiner:1988,Turbiner:2016},
\begin{eqnarray}
\label{sl4R}
 {\cal J}_i^- \ &=& \ \frac{\pa}{\pa \lambda_i}\ ,\qquad \quad i=1,2,\dots,6\ , \non  \\
 {{\cal J}_{ij}}^0 \ &=& \
               \lambda_i \frac{\pa}{\pa \lambda_j}\ , \qquad i,j=1,2,3\dots,6 \ , \\
 {\cal J}^0(N)\ &=&  \ \sum_{i=1}^{6} \lambda_i \frac{\pa}{\pa \lambda_i}-N\, , \non \\
 {\cal J}_i^+(N) \ &=& \ \lambda_i \,{\cal J}^0(N)\ =\
    \lambda_i\, \left( \sum_{j=1}^{6} \lambda_j\frac{\pa}{\pa \lambda_j}-N \right)\ ,
       \quad i=1,2,\dots,6\ ,
\end{eqnarray}
where $N$ is parameter and
\[
 \lambda_1\equiv\rho_{12}\ ,\qquad \lambda_2\equiv\rho_{13}\ , \qquad \lambda_3\equiv\rho_{14} \qquad \lambda_4\equiv\rho_{23}\ , \qquad \lambda_5\equiv\rho_{24}\qquad \lambda_6\equiv\rho_{34}\ .
\]
If $N$ is non-negative integer, a finite-dimensional representation space occurs,
\begin{equation}
\label{P3}
     {\cal P}_{N}\ =\ \langle  \lambda_1^{p_1}\, \lambda_2^{p_2}\, \lambda_3^{p_3}\,\lambda_4^{p_4} \,\lambda_5^{p_5} \,\lambda_6^{p_6} \vert \ 0 \le p_1+p_2+p_3+p_4+p_5+p_6 \le N \rangle\ .
\end{equation}
Explicitly, the operator (\ref{addition3-3rho})
\begin{equation}
\label{HRex}
  \frac{1}{2}\, \De_{radial}({\cal J}) \ = \ 2(\, {\cal J}_{11}^0\,{\cal J}_1^- + {\cal J}_{22}^0\,{\cal J}_2^-
      + {\cal J}_{33}^0\,{\cal J}_3^-  + {\cal J}_{44}^0\,{\cal J}_4^-  + {\cal J}_{55}^0\,{\cal J}_5^-  + {\cal J}_{66}^0\,{\cal J}_6^-  \,)\
\end{equation}
\[
+\ d \,({\cal J}_1^- + {\cal J}_2^- + {\cal J}_3^- + {\cal J}_4^-  + {\cal J}_5^-  + {\cal J}_6^-)\
\]
\[
+  \bigg[{\cal J}_{11}^0\,({\cal J}_2^- + {\cal J}_3^-  + {\cal J}_4^-  + {\cal J}_5^- ) + {\cal J}_{22}^0\,({\cal J}_1^- + {\cal J}_3^-  + {\cal J}_4^-  + {\cal J}_6^- )
\]
\[
+  {\cal J}_{33}^0\,({\cal J}_1^- + {\cal J}_2^-  + {\cal J}_5^-  + {\cal J}_6^- ) + {\cal J}_{44}^0\,({\cal J}_1^- + {\cal J}_2^-  + {\cal J}_5^-  + {\cal J}_6^- )
\]
\[
+ {\cal J}_{55}^0\,({\cal J}_1^- + {\cal J}_3^-  + {\cal J}_4^-  + {\cal J}_6^- ) +  {\cal J}_{66}^0\,({\cal J}_2^- + {\cal J}_3^-  + {\cal J}_4^-  + {\cal J}_5^- ) \bigg]
\]
\[
-2 \bigg[ {\cal J}_{12}^0\,{\cal J}_4^-  +  {\cal J}_{13}^0\,{\cal J}_5^-   +  {\cal J}_{21}^0\,{\cal J}_4^-   +   {\cal J}_{23}^0\,{\cal J}_6^-   +  {\cal J}_{31}^0\,{\cal J}_5^-   +   {\cal J}_{32}^0\,{\cal J}_6^-
\]
\[
 +   {\cal J}_{41}^0\,{\cal J}_2^-   +  {\cal J}_{45}^0\,{\cal J}_6^- + {\cal J}_{54}^0\,{\cal J}_6^-  +  {\cal J}_{62}^0\,{\cal J}_3^-  +   {\cal J}_{64}^0\,{\cal J}_5^-  +   {\cal J}_{51}^0\,{\cal J}_3^-        \bigg] \ .
\]
It acts on (\ref{P3}) as filtration.

\subsection{The Laplace Beltrami operator, underlying geometry}

The remarkable property of the algebraic operator ${\De_{radial}}(\rho)$ (\ref{addition3-3rho}) is its gauge-equivalence to the Schr\"odinger operator: there exists the gauge factor $\Gamma$ such that
\begin{equation}
\label{DLB4}
     \Gamma^{-1}\,\Delta_{{radial}}(\rho)\, \Gamma\ =\ {\Delta_{LB}}(\rho) - {V}_{eff}(\rho)\ ,
\end{equation}
where $\De_{LB}$ is the Laplace-Beltrami operator

\begin{equation}
\label{LBop}
{\Delta_{LB}}(\rho) \ = \ \sqrt{D(\rho)}\,\pa_\mu \frac{1}{\sqrt{D(\rho)}}g^{\mu \nu}\,\pa_\nu \ , \qquad \pa_\nu \equiv \frac{\pa}{\pa \rho_\nu} \ ,
\end{equation}
see (\ref{gmn33-rho}), is given by
\begin{equation}
\label{GamFac}
  \Gamma \ =\ (F_1 \,F_2)^{-1/4}({V}_4^2)^{1-d/4}\ =\ F_1^{\frac{3 - d}{4}}\,F_2^{-1/4} \ ,
\end{equation}
see (\ref{determinant4}), (\ref{F1}), (\ref{F2}) and the \emph{effective} potential is
\begin{equation}
\label{VeffF1}
V_{eff} \ =\  \frac{3\,\tilde { V}_2^4
+112\,\tilde{ V}_3^2}{32\,F_2} \ + \ \frac{(d-5)(d-3)\tilde {V}_3^2}{72\,F_1} \ .
\end{equation}

Therefore $\Gamma$ is of geometric nature: it can be rewritten in terms of volumes (equivalently, volume variables, see below). The effective potential becomes singular at boundary of the configuration space.

Eventually, taking into account the gauge rotation (\ref{DLB4}) we arrive at the six-dimensional Hamiltonian
\begin{equation}
\label{H-3-3r-r}
    {\cal H}_{LB} (r) \ = \ -\De_{LB}(r)\  + \   V(r) \  + \   V_{eff}(r)   \ ,
\end{equation}
in the space of $r$-relative distances, or
\begin{equation}
\label{H-3-3r-rho}
    {\cal H}_{LB} (\rho) \ =\ -\De_{LB}(\rho) \  + \  V(\rho) \  + \   V_{eff}(\rho)\ ,
\end{equation}
in $\rho$-space. The Hamiltonians (\ref{H-3-3r-r}) and (\ref{H-3-3r-rho}) describe the six-dimensional quantum particle moving in the curved space with metric $g^{\mu \nu}$ and kinetic energy $\De_{LB}$, in particular, in $\rho$-space with metric $g^{\mu \nu}(\rho)$ (\ref{gmn33-rho}) and kinetic energy $\De_{LB}(\rho)$.

Making the de-quantization of (\ref{H-3-3r-rho}) we arrive at a six-dimensional classical system which is characterized by the Hamiltonian,
\begin{equation}
\label{H-3-3r-rho-class}
    {\cal H}_{LB}^{(c)} (\rho) \ =\ g^{\mu \nu}(\rho)\,P_{\mu}\, P_{\nu} \  + \  V(\rho)  \ + \  V_{eff}(\rho)\ ,
\end{equation}
where $P_{\mu}\,, \ \mu=12,13,14,23,24,34$ is classical canonical momenta in $\rho$-space and $g^{\mu \nu}(\rho)$ is given by (\ref{gmn33-rho}). This operator (\ref{H-3-3r-rho-class}) is suitable to investigate special configurations (trajectories) for the classical four-body system. It is worth to mention that even in the planar case, the dynamics of the classical four-body problem is very rich \cite{Erdi}-\cite{Albouy}.


\section{Reduction to lower dimensions: $d=1,2$}

At $d=2$ (planar systems) and $d=1$ (a system on the line) the number of independent $\rho$-variables reduces from six to five and three, respectively, and the expression (\ref{addition3-3rho}) for the operator $\Delta_{radial}$ ceases to be valid. In particular, the determinant of the metric defined by the coefficients of the second order derivatives in (\ref{addition3-3rho}) vanishes, the operator $\Delta_{radial}$ is not invertible. This makes the cases $d=2$ and $d=1$ quite distinct from $d\geq3$.

In particular, one can ask the question: \emph{do exist variables for which $\De_{radial}$ is an algebraic operator at $d=2$?}. In this section we provide a partial answer to this question. To this end, in addition to the $\rho$-representation we will introduce two new representations in terms of purely geometric variables (see below) obtained from the \emph{tetrahedron of interaction}. We call them \emph{volume}-variables and $u$-variables, respectively. More importantly, the volume-representation can be easily extended to the general $n$-body case.

\subsection{Volume variables representation}

Let us consider, assuming $d\geq3$, the following change of variables

\[
(\,\rho _{12},\,\rho _{13},\,\rho _{14},\,\rho _{23},\,\rho _{24},\,\rho _{34},\,) \  \Rightarrow \ (\ {\cal V},\ S,\ P,\ q_1,\ q_2,\ q_3\ ) \ ,
\]
where
\begin{equation}
\label{V4}
\begin{aligned}
 {\cal V}  \ \equiv & \  V_4^2  \  = \  \frac{1}{144} \bigg[\,\left[\left(\rho _{13}+\rho _{14}+\rho _{23}+
  \rho _{24}\right) \rho _{34}-\left(\rho _{13}-\rho _{14}\right) \left(\rho _{23}-\rho _{24}\right)-
  \rho _{34}^2 \right] \rho _{12}
\\ & - \ \rho _{13}^2 \rho _{24} \ - \ \rho _{34} \rho _{12}^2 \ +
\  \rho _{23} \left[\left(\rho _{14}-\rho _{24}\right) \rho _{34}-\rho _{14} \left(\rho _{14}+
  \rho _{23}-\rho _{24}\right)\right]
\\ &  \ + \ \rho _{13} \left[\,\rho _{14} \left(\rho _{23}+\rho _{24}-
  \rho _{34}\right)+\rho _{24} \left(\rho _{23}-\rho _{24}+\rho _{34}\right)\right]\bigg]  \ ,
\end{aligned}
\end{equation}
is the square of the volume of the \emph{tetrahedron of interaction}, the variable
\begin{equation}
\label{VarS}
S \ \equiv \ \tilde {V}_3^2 \ = \ S_{1} \ + \ S_{2} \ + \ S_{3} \ + \ S_{4} \ ,
\end{equation}
is nothing but the sum of the areas squared of its four faces (see (\ref{CFrho})),
and the variable
\begin{equation}
\label{VarP}
   P  \ \equiv \ \tilde {V}_2^2 \ = \  \rho _{12} \ + \ \rho_{13} \ + \ \rho_{14} \ + \  \rho_{23} \ + \ \rho_{24} \ + \
   \rho _{34} \ ,
\end{equation}
is just the sum of all the six edges (squared). This variable is nothing but the square of hyper-radius in the space of relative motion or, in other words, in the space of relative distances.

The nature of these three variables $({\cal V},S,P)$ is purely geometric, they are homogeneous polynomials in $\rho$-variables of dimension three, two and one, respectively. Notice that these quantities define the effective potential $V_{eff}$ (\ref{VeffF1}). We call them \emph{volume variables}.

Clearly, ${\cal V},\,S$ and $P$ are $S_4$-invariant under the permutations of the four body positions (vertices of the tetrahedron). However, only the variable $P$ is $S_6$ invariant under 
the permutations of the six $\rho$-variables (edges of the tetrahedron).
The remaining three variables
\[
(\ q_1,\ q_2,\ q_3 \ ) \ ,
\]
can be chosen as $q_1=\rho_{12}$, $q_2=\rho_{13}$ and $q_3=\rho_{14}$, $d>1$. The specific form of the $q$-variables is irrelevant for our purposes,
see below.

In the above mentioned variables, $\Delta_{radial}$ (\ref{addition3-3rho}) can be  further decomposed
into the sum of two operators
\begin{equation}
\label{DEsum}
\Delta_{radial} \ = \ \Delta_{g} \ + \   \Delta_{q,g} \ ,
\end{equation}
with the following properties:
\begin{itemize}
  \item $\Delta_{g}=\Delta_{g}({\cal V},S,P)$: it is an algebraic operator for any $d$, it only depends on \emph{volume variables} ${\cal V},\,S$, and $P$ and its derivatives,
\begin{equation}
\begin{aligned}
{\label{DEg}}
{\De}_g  \ = & \   \frac{2}{9}\,{\cal V}\,S\,\pa^2_{{\cal V},{\cal V}}  \ +  \  \big(  54\,{\cal V} \, + \, \frac{1}{2}S\,P \  \big)\,\pa^2_{S,S} \ + \ 8\,P\,\pa^2_{P,P}
\\ &  +  \  32\,S\,\pa^2_{S,P} \  +  \ 2\,{\cal V}\,\big(    P\,\pa^2_{{\cal V},S} \ + \ 24\,\pa^2_{{\cal V},P} \  \big) \ + \ \frac{1}{9}(d-2)\,S\,\pa_{{\cal V}}
\\ &  + \  \frac{1}{2}(d-1)\,P\,\pa_{S} \ + \ 12\,d\,\pa_{P}     \ .
\end{aligned}
\end{equation}

  \item $\Delta_{q,g}=\Delta_{q,g}({\cal V},S,P,q_1,q_2,q_3)$: for arbitrary $d$, this operator annihilates any volume-variables dependable function, namely $\Delta_{q,g}\,f({\cal V},S,P)\ =\ 0$\,. We do not give its explicit form.
  \item $[\Delta_{g},\,\Delta_{q,g}] \ \neq 0 $\ .
\end{itemize}

The operator (\ref{DEg}) is $sl(4, {\bf R})$-Lie-algebraic, see e.g. \cite{Turbiner:2016}, and it is gauge-equivalent to a three-dimensional Schr\"odinger operator in curved space with $d$-independent metric (see below). For this operator ${\De}_g$, the reduction from $d \geq 3$ to $d=2$ simply corresponds to imposing the condition ${\cal V}=0$ together with $d=2$. In turn, the reduction to $d=1$ occurs when two conditions are imposed ${\cal V}=S=0$ together with $d=1$. Both limits to $d=2$ and $d=1$ are geometrically transparent and, more importantly, ${\De}_g$ remains algebraic.

The $d$-independent metric of ${\De}_g$ is given by
\begin{equation}
\label{gVSP}
g^{\mu \nu}({\cal V},S,P) \ = \ \left(
\begin{array}{ccc}
       \frac{2}{9}\,{\cal V}\,S & {\cal V}\,P & 24\,{\cal V} \\
       {\cal V}\,P & 54\,{\cal V} \, + \, \frac{1}{2}S\,P & 16\,S \\
       24\,{\cal V} & 16\,S & 8\,P \\
\end{array}
\right) \ ,
\end{equation}
and its determinant factorizes
\[
D_g({\cal V},S,P) \ \equiv \ {\text {Det}} \,g^{\mu\,\nu} \ = \ G_1\,G_2 \ ,
\]
where
\[
G_1 \ = \ \frac{8}{9}\, {\cal V} \ ,
\]
\[
 G_2 \ = \ S^2 \left(P^2-64\, S\right) \ - \ 9\, P\, {\cal V}\,
 \left(P^2-72 \,S\right) \ - \ 34992\, {\cal V}^2 \ .
\]

The boundary of the configuration space is defined by ${\cal V}=0$. Using the gauge factor
\begin{equation}
\label{GammaVSP}
\Gamma_g \ = \  G_1^{\frac{3-d}{4}} \, G_2^{-\frac{1}{4}} \ ,
\end{equation}
to make gauge rotation of the operator $\Delta_g$ we arrive at the Schr\"odinger operator
\begin{equation}
\label{}
     \Gamma_g^{-1}\,{\De_g}\, \Gamma_g\ =\ {\De_{LB}}({\cal V},S,P) \ - \  \tilde V_g \ ,
\end{equation}
with effective potential
\[
  \tilde V_g \ = \ (d-5) (d-3)\frac{ S}{81 \,G_1}\ + \
 \frac{\left(P^2-48\, S\right) (324\, {\cal V}-P \,S)}{8 \,G_2} \ ,
\]
where the first term vanishes at $d=3,5$, and $\De_{LB}$ is the Laplace-Beltrami operator
\begin{equation}
\label{LB-V}
{\Delta_{LB}}({\cal V},S,P) \ = \ \sqrt{D_g}\,\partial_\mu \frac{1}{\sqrt{D_g}}g^{\mu \nu}\,\partial_\nu \ ,
\end{equation}
here $\nu,\,\mu$ label the variables ${\cal V},S,P$, and $g^{\mu \nu}$ is given by (\ref{gVSP}).

Eventually, for the original four-body problem (\ref{Hrel-Mod}) in the space of relative motion, provided that the potential only depends on the volume variables and taking into account the gauge rotation $\Gamma_g$ (\ref{GammaVSP}), we arrive at the gauge-equivalent Hamiltonian
\begin{equation}
\label{HLBg}
    {\cal H}_{LB} ({\cal V},S,P) \ =\ -\De_{LB}({\cal V},S,P)\  + \  \tilde V_g({\cal V},S,P) \  + \   V({\cal V},S,P)   \ ,
\end{equation}
in the space of volume variables. The Hamiltonian (\ref{HLBg}) describes a three-dimensional quantum particle moving in the curved space parametrized by ${\cal V},S,P$ with metric $g^{\mu \nu}$ (\ref{gVSP}) and kinetic energy $\De_{LB}$. The form of (\ref{HLBg}) implies the possible existence of a subfamily of eigenfunctions in the form of a multiplicative factor times an inhomogeneous polynomial in the variables $({\cal V},S,P)$. The volume variables can be generalized to the case of non equal masses (see Appendix B).

\bigskip

\subsubsection{Towards $d=2$}

\bigskip

Let us assume that $V=V({\cal V},S,P)$ in (\ref{Hrel-Mod}). In this case, we can ignore the operator $\Delta_{q,g}$ in (\ref{DEsum}). Now, for $d=2$ the volume of the tetrahedron of interaction vanishes identically ${\cal V}=0$. Therefore, the operator $\Delta_{g}$ (\ref{DEg}) reduces to

\begin{equation}
\begin{aligned}
{\De}_g|_{d=2}  \ = & \   \frac{1}{2}S\,P\,\pa^2_{S,S} \ + \ 8\,P\,\pa^2_{P,P}\ +  \  32\,S\,\pa^2_{S,P} \ + \  \frac{1}{2}\,P\,\pa_{S} \ + \ 24\,\pa_{P}     \ .
\end{aligned}
\end{equation}
Thus, in the limit $d\rightarrow 2$, ${\De}_g$ remains algebraic (more precisely $sl(3, {\bf R})$-Lie-algebraic). The corresponding metric of ${\De}_g|_{d=2}$ takes the form
\begin{equation}\label{gmuSP}
g^{\mu \nu}(S,P) \ = \ \left(
\begin{array}{cc}
         \frac{1}{2}S\,P & 16\,S \\
        16\,S & 8\,P \\
\end{array}
\right) \ ,
\end{equation}
and its determinant factorizes as
\[
D_k(S,P) \ \equiv \ {\text {Det}} \,g^{\mu\,\nu} \ = \ K_1\,K_2 \ ,
\]
where
\[
K_1 \ = \ S \ ,
\]
\[
K_2 \ = \ P^2  -  64\,S \ .
\]
The boundary of configuration space is defined by $S=0$. Using the gauge factor
\[
\Gamma_{gk} \ = \  (K_1\,K_2)^{-\frac{1}{4}} \ ,
\]
to gauge-rotate the restricted operator ${\De}_g|_{d=2}$ we obtain
\begin{equation}
\label{Gammagk}
     \Gamma_{gk}^{-1}\,{\De}_g|_{d=2}\, \Gamma_{gk}\ =\ {\De_{LB}}(S,P) \ - \  \tilde V_{gk} \ ,
\end{equation}
here the effective potential reads
\[
\tilde V_{gk} \ = \ \frac{P^3}{32 \,S \,\left(P^2-64 \,S\right)} \ .
\]

Eventually, for the original four-body problem (\ref{Hrel-Mod}) in the space of relative motion, provided that the potential only depends on the two volume variables $(S,P)$ and taking into account the gauge rotation $\Gamma_{gk}$ (\ref{Gammagk}), we arrive at the gauge-equivalent two-dimensional Hamiltonian
\begin{equation}
\label{HLBg2}
    {\cal H}_{LB} (S,P) \ =\ -\De_{LB}(S,P)\  + \  \tilde V_{gk}(S,P) \  + \   V(S,P)   \ .
\end{equation}
The Hamiltonian (\ref{HLBg2}) describes a two-dimensional quantum particle moving in the curved space with metric $g^{\mu \nu}$ (\ref{gmuSP}). The form of (\ref{HLBg2}) suggests the possible existence of a subfamily of eigenfunctions in the form of a multiplicative factor times a polynomial in the variables $(S,P)$.

\subsubsection{Towards $d=1$}

For $d=1$, both the volume variable ${\cal V}$ and the area variable $S$ vanish identically. In this case the algebraic operator (\ref{DEg}) depends on the variable $P$ alone, it has the form

\begin{equation}
\begin{aligned}
{\De}_g|_{d=1}  \ = & \   8\,P\,\pa^2_{P,P}\  + \ 12\,\pa_{P}     \ ,
\end{aligned}
\end{equation}
which after a suitable gauge rotation and upon the addition of an harmonic potential $\sim \omega^2\,P$ it becomes the Laguerre operator.
We again point out that the form of the operator ${\De}_g|_{d=1}$ implies the existence of a subfamily of solutions of the original four-body problem (\ref{Hrel-Mod}) in the space of relative motion, which only depend on the variable $P$\,.

Let us clarify the space "degeneration" from $d=3$ to $d=1$ with a concrete example. For $d=1$, the number of functionally independent variables (degrees of freedom) in $\Delta_{radial}$ (\ref{DEsum}) is three and the first operator $\Delta_{g}$ in (\ref{DEsum}) solely depends on the variable $P$, see (59). Therefore, the operator $\Delta_{q,g}$ must involve two $q$-variables only.

Now, without loss of generality let us choose
\[
q_1 \ = \ \rho_{12} \ , \qquad q_2 \ = \ \rho_{23} \ , \qquad q_3 \ = \ \sqrt{\rho_{12}} + \sqrt{\rho_{23}}-\sqrt{\rho_{13}}\ ,
\]
as the $q$-variables for $d > 1$. For $d=1$, $\rho_{ij}\equiv{(x_i-x_j)}^2$ and
$\infty > x_1 > x_2 > x_3 > x_4 > 0$, the variable $q_3$ vanishes identically and $$P=3(\rho_{12}+\rho_{34})+4\,\sqrt{\rho_{23}}\,(\sqrt{\rho_{12}}+\sqrt{\rho_{23}}+\sqrt{\rho_{34}}) + 2\,\sqrt{\rho_{12}}\,\sqrt{\rho_{34}}\ .$$

For $d \geq 2$, the operator $\Delta_{q,g}$ (\ref{DEsum}) reads
\begin{equation}
\label{q123}
\begin{aligned}
& \Delta_{q,g} \ = \  4\,q_1\,\pa^2_{q_1,q_1} \ + \  4\,q_2\,\pa^2_{q_2,q_2} \ - \ 2 \left(q_3^2-2 \sqrt{q_1} q_3-2 \sqrt{q_2} q_3+2 \sqrt{q_1} \sqrt{q_2}\right)\pa^2_{q_1,q_2}
\\ &
\ + \ \frac{q_3 \left(q_3^2-3 \sqrt{q_1} q_3-4 \sqrt{q_2} q_3+4 \sqrt{q_1} \sqrt{q_2}+2 q_1+4 q_2\right)}{\sqrt{q_2} \left(\sqrt{q_1}+\sqrt{q_2}-q_3\right)}\pa^2_{q_1,q_3} \ + \ 2\,d\,(\pa_{q_1}\,+\,\pa_{q_2})
\\ &
\ + \ \frac{q_3 \left(q_3^2-4 \sqrt{q_1} q_3-3 \sqrt{q_2} q_3+4 \sqrt{q_1} \sqrt{q_2}+4 q_1+2 q_2\right)}{\sqrt{q_1} \left(\sqrt{q_1}+\sqrt{q_2}-q_3\right)}\pa^2_{q_2,q_3}
\\ &
\ + \ \frac{(d-1) \left(\sqrt{q_1} \sqrt{q_2}-q_3 \sqrt{q_2}+q_1+q_2-\sqrt{q_1} q_3\right)}{\sqrt{q_1} \sqrt{q_2} \left(\sqrt{q_1}+\sqrt{q_2}-q_3\right)}\pa_{q_3}
\\ &
\ + \ 8\,{\cal V}\,\bigg[\pa^2_{{\cal V},q_1} \, + \, \pa^2_{{\cal V},q_2}\ + \ \frac{\sqrt{q_1} \sqrt{q_2}-q_3 \sqrt{q_2}+q_1+q_2-\sqrt{q_1} q_3}{32\, \sqrt{q_1}\, \sqrt{q_2} \left(\sqrt{q_1}+\sqrt{q_2}-q_3\right)}\,\pa^2_{{\cal V},q_3}\bigg]
\\ &
\ + \ \sum_{i=1}^{3} A_i\, \pa^2_{S,q_i}\ \ + \ 8\,(2\,q_1\,\pa^2_{P,q_1}\,+\,2\,q_2\,\pa^2_{P,q_2} \, +\, q_3\,\pa^2_{P,q_3}) \ ,
\end{aligned}
\end{equation}
where the coefficients $A_i$ are functions of (${\cal V},S,P,q_1,q_2,q_3$).
At $d=1$ they vanish: $A_i=0$. Also all terms involving derivatives $\pa_{\cal V}$ vanish.
Eventually, in the limit $d \rightarrow 1$ we end up with
\begin{equation}
\label{}
\begin{aligned}
 \Delta_{q,g}\mid_{d=1} \ & = \  4\,q_1\,\pa^2_{q_1,q_1} \ + \  4\,q_2\,\pa^2_{q_2,q_2} \ - \ 4 \,\sqrt{q_1} \sqrt{q_2}\,\pa^2_{q_1,q_2}\ + \ 2\,(\pa_{q_1}\,+\,\pa_{q_2})
\\ &  \ + \ 16\,(\,q_1\,\pa^2_{P,q_1}\,+\,q_2\,\pa^2_{P,q_2}\,) \ ,
\end{aligned}
\end{equation}
thus, the operator $\Delta_{q,g}$ depends on two $q$-variables only.
Finally, for $\Delta_{radial}$ we arrive to the well defined three-dimensional operator
\begin{equation}
\label{Delta-d1}
\begin{aligned}
 \De_{radial}\mid_{d=1} \ & =  \De_{g}\mid_{d=1} \ + \ \De_{q,g}\mid_{d=1}
\\ &
  =  \  8\,P\,\pa^2_{P,P}\  + \ 12\,\pa_{P} \ +\  4\,q_1\,\pa^2_{q_1,q_1} \ + \  4\,q_2\,\pa^2_{q_2,q_2}
\\ &
 \ - \ 4 \,\sqrt{q_1} \sqrt{q_2}\,\pa^2_{q_1,q_2}\ + \ 2\,(\pa_{q_1}\,+\,\pa_{q_2}) \ + \ 16\,(\,q_1\,\pa^2_{P,q_1}\,+\,q_2\,\pa^2_{P,q_2}\,)\ ,
\end{aligned}
\end{equation}
which after a suitable change of variables is reduced to the algebraic operator (\ref{Drel3-1}).

\subsection{u-variables representation}

It is worth mentioning another decomposition of the operator $\Delta_{radial}$ (\ref{addition3-3rho}), assuming $d \geq 3$, in the variables
\begin{equation}
\label{uq-d3}
(\,\rho _{12},\,\rho _{13},\,\rho _{14},\,\rho _{23},\,\rho _{24},\,\rho _{34},\,) \ \Rightarrow \ (\ u_1,\,u_2,\,u_3,\,q_1,\,q_2,\,q_3   \ )\ ,
\end{equation}
where
\[
   u_1 = \rho_{12} + \rho_{34}  \ ; \qquad  u_2 = \rho_{13}+\rho_{24} \ ; \qquad
   u_3 = \rho_{23}+\rho_{14}\ ,
\]
are nothing but the sum of two disconnected edges (squared), thus, without common vertices, of the \emph{tetrahedron of interaction}; they are geometrical-type variables.  They also are $S_4$ invariant under the permutations of the four body positions.

For simplicity we can choose $q_1=\rho_{12}$, $q_2=\rho_{13}$ and $q_3=\rho_{14}$, ($d\geq3$).
In the new variables (\ref{uq-d3}), the operator $\De_{radial}$ (\ref{addition3-3rho}) is decomposed
in the sum of two operators
\begin{equation}
\label{DRG}
\Delta_{radial} \ = \  \Delta_u \ + \  \Delta_{q,u} \ ,
\end{equation}
with the following properties:
\begin{itemize}
  \item $\Delta_u=\Delta_u(u_1,u_2,u_3)$: it is an algebraic operator for any $d$ and involves the $u$-variables and its derivatives only
  \begin{equation}
  \begin{aligned}
  \label{OPDu}
    & \frac{1}{2}{\De}_u  \ = \ 2\,u_1\,\pa^2_{u_1,u_1} \ + \ 2\,u_2\,\pa^2_{u_2,u_2} \ + \ 2\,u_3\,\pa^2_{u_3,u_3}
\\
    & + \ 2\,(u_1+u_2-u_3)\pa^2_{u_1,u_2}\ + \ 2\,(u_1+u_3-u_2)\pa^2_{u_1,u_3}\ + \ 2\,(u_2+u_3-u_1)\pa^2_{u_2,u_3}
\\
    & + \ 2\,d\,(\pa_{u_1} + \pa_{u_2}+ \pa_{u_3})     \ .
\end{aligned}
\end{equation}

  \item $\Delta_{q,u}=\Delta_{q,u}(u_1,u_2,u_3,q_1,q_2,q_3)$: for any $d$, it annihilates any $u$-dependable function, namely $\Delta_{q,u}\,f(u_1,u_2,u_3)=0$ \ .

  \item $[\Delta_{u},\,\Delta_{q,u}] \ \neq 0 $\ .

\end{itemize}

If the original 4-body potential depends on $u$-variables only the decomposition (\ref{DRG}) implies
the further reduction of the already reduced spectral problem (\ref{Hrel-Mod}) to
\[
   (-\Delta_u + V (u)) \Psi(u)\ =\ E \Psi(u)\ .
\]
The operator $\Delta_u$ (\ref{OPDu}) is $sl(4, {\bf R})$-Lie-algebraic with a flat $d$-independent metric
\begin{equation}\label{gu}
g^{\mu \nu}(u) \ = \ \left(
\begin{array}{ccc}
       4\,u_1 & 2\,(u_1+u_2-u_3) & 2\,(u_1+u_3-u_2) \\
       2\,(u_1+u_2-u_3) & 4\,u_2 & 2\,(u_2+u_3-u_1) \\
       2\,(u_1+u_3-u_2) & 2\,(u_2+u_3-u_1) & 4\,u_3 \\
\end{array}
\right) \ ,
\end{equation}
with rather simple factorizable expression for its determinant
\[
D(u) \ \equiv \ {\text {Det}} \,g^{\mu\,\nu}(u) \ = \ 32\,(u_1+u_2-u_3)(u_1+u_3-u_2)(u_2+u_3-u_1) \ .
\]
The boundary of the configuration space is defined by $D(u)=0$.
Moreover, using the gauge factor
\[
\Gamma_u \ = \  D(u)^{\frac{1-d}{4}} \ ,
\]
for gauge rotation of the operator $\Delta_u$ we obtain a gauge-equivalent three-dimensional Schr\"odinger operator
\begin{equation}
\label{GammaU}
     \Gamma_u^{-1}\,{\De_u}\,(u)\, \Gamma_u\ =\ {\De_{LB}}(u) \ - \  \tilde V_u(u) \ ,
\end{equation}
with the effective potential of the form
\[
\tilde V_u(u) \ = \ (d-1)(d-3)\frac{ \left(u_1^2 + u_2^2 + u_3^2 -2\,(u_1\,u_2 + u_1\,u_3 +u_2\,u_3 ) \right)}{2 \left(u_1-u_2-u_3\right) \left(u_1+u_2-u_3\right) \left(u_1-u_2+u_3\right)} \ .
\]

Finally, for the original four-body problem (\ref{Hrel-Mod}) in the space of relative motion, provided that the potential only depends on the $u$-variables, taking into account the gauge rotation $\Gamma_u$ (\ref{GammaU}) and assuming the $u$-dependent solutions are studied only,
we arrive at the gauge-equivalent three-dimensional Hamiltonian
\begin{equation}
\label{HLBu}
    {\cal H}_{LB} (u_1,u_2,u_3) \ =\ -\De_{LB}(u_1,u_2,u_3)\  + \  \tilde V_u(u_1,u_2,u_3) \  + \   V(u_1,u_2,u_3)   \ ,
\end{equation}
in the space of $u$-variables. The Hamiltonian (\ref{HLBu}) also describes a three-dimensional quantum particle moving in the flat space parametrized by $u_1,u_2,u_3$ with metric $g^{\mu \nu}$ (\ref{gu}) and kinetic energy $\De_{LB}(u)$. The form of (\ref{HLBu}) implies the possible existence of a subfamily of eigenfunctions in the form of a multiplicative factor times an inhomogeneous polynomial in the variables $(u_1,u_2,u_3)$. The $u$-variables do not admit a generalization to the case of non equal masses.

\subsubsection{Towards $d=2$ and $d=1$}

Unlike the volume variables ${\cal V}$ and $S$, the $u$-variables (\ref{uq-d3}) are not subject to any constraint at $d=2$ (${\cal V}=0$) and $d=1$ (${\cal V}=S=0$). Moreover, for the operator ${\De_u}$ (\ref{OPDu}) the passage to lower dimensions is non-singular. Coming to $d=1,2$ the overall multiplicative factor in front of the first derivative terms in (\ref{OPDu}) changes only.

However, for $d=2$ the number of variables (degrees of freedom) in $\Delta_{radial}$ (\ref{DEsum}) is reduced to five. Therefore, in this case the operator $\Delta_{q,u}$ in (\ref{DRG}) must involve two $q$-variables only
\[
\Delta_{q,u}|_{d=2} \ = \ \sum_{i+j=1}^{2} Y_{i,j}\,\partial^i_{q_1}\pa^j_{q_2} \ ,
\]
with coefficients $Y_{i,j}=Y_{i,j}(u_1,u_2,u_3,q_1,q_2)$. In general, $\Delta_{q,u}|_{d=2}$ is not an algebraic operator.

At $d=1$ the operator $\Delta_{q,u}$ vanishes, $\Delta_{q,u}=0$, while the algebraic operator $\Delta_u$ (\ref{OPDu}) after a suitable gauge rotation and change of variables describes the kinetic energy of relative motion of the four-body ($A_3$) rational Calogero model with potential (\ref{VA3}), see \cite{RT:1996}.

\subsection{P-variable representation}

Let us pay attention that in (\ref{DEg}), the coefficients in front of the second and the first derivative in $P$ (\ref{VarP}) do not involve the volume variables $\cal V$ (\ref{V4}) and $S$ (\ref{VarS}). Furthermore, in reality the variable
\[
   P \ = \ u_1 \ + \ u_2 \ + \ u_3 \ = \ \rho_{12} + \rho_{13} + \rho_{14}+\rho_{23} + \rho_{24}+\rho_{34}
   \ .
\]
is nothing but the sum of the $u$-variables (\ref{uq-d3}), which appear at the algebraic operator $\Delta_u$ (\ref{OPDu}) at any $d$. Based on these two facts let us make, assuming $d \geq 3$, change of variables
\begin{equation}
\label{Pd3}
(\,\rho _{12},\,\rho _{13},\,\rho _{14},\,\rho _{23},\,\rho _{24},\,\rho _{34},\,) \ \Rightarrow \ (\ P,\,q_1,\,q_2,\,q_3,\,q_4,\,q_5   \ )\ .
\end{equation}
We call it $P$-representation. It is worth to note that $P$ is, up to an overall constant factor, the unique linear combination of $\rho$-variables that is both $S_4$-invariant under the permutations of the four body positions as well as $S_6$-invariant under the permutations of the six $\rho$'s. In the same time the $q$-variables form a set of well-defined quantities such that the Jacobian of the transformation (\ref{Pd3}) is not singular.

In the variables (\ref{Pd3}), the operator $\De_{radial}$ (\ref{addition3-3rho}) admits the decomposition
\begin{equation}
\label{DRP}
\Delta_{radial} \ = \  \Delta_P \ + \  \Delta_{q,P} \ ,
\end{equation}
with the following properties:
\begin{itemize}
  \item $\Delta_P=\Delta_P(P)$: it is an algebraic operator for any $d$ and involves the $P$-variable and its derivatives only
  \begin{equation}
  \begin{aligned}
  \label{OPDP}
{\De}_P  \ = & \    8\,P\,\pa^2_{P,P} \ + \ 12\,d\,\pa_{P}     \ .
\end{aligned}
\end{equation}

  \item $\Delta_{q,P}=\Delta_{q,P}(P,q_1,q_2,q_3,q_4,q_5)$: for any $d$, it annihilates any $P$-dependable function, namely $\Delta_{q,P}\,f(P)=0$ \ .

  \item $[\Delta_{P},\,\Delta_{q,P}] \ \neq 0 $\ .

\end{itemize}
Using a gauge factor
\[
\Gamma_P \ = \  P^{\frac{1-3\,d}{4}} \ ,
\]
for gauge rotation of the operator $\Delta_P$ we obtain the gauge-equivalent, one-dimensional Schr\"odinger operator
\begin{equation}
\label{GammaP}
     \Gamma_P^{-1}\,{\De_P}\,(P)\, \Gamma_P\ =\ {\De_{LB}}(P) \ - \  \tilde V_P(P) \ ,
\end{equation}
with the Laplace-Beltrami operator
\[
{\De_{LB}}(P)  \ = \  4\,(\,2\,P\,\pa^2_{P,P} \ + \ \pa_{P}\,)\ ,
\]
with metric
\[
g^{11} \ =\ 8\,P \ ,
\]
and an effective potential of the form
\[
\tilde V_P(P) \ = \ \frac{3 \,(d-1) (3 \,d-1) }{2\, P} \ .
\]

In conclusion, for the original four-body problem (\ref{Hrel-Mod}) in the space of relative motion, provided that the potential depends on the $P$-variable only and taking into account the gauge rotation $\Gamma_P$ (\ref{GammaP}), we arrive at the gauge-equivalent one-dimensional Hamiltonian
\begin{equation}
\label{HLBP}
    {\cal H}_{LB} (P) \ =\ -\De_{LB}(P)\  + \  \tilde V_P(P) \  + \   V(P)   \ .
\end{equation}
The form of (\ref{HLBP}) implies the possible existence of a subfamily of eigenfunctions in the form of a $P$-dependent multiplicative factor times an inhomogeneous polynomial in $P$. For $d=1$, this remarkable property was previously pointed out in \cite{MMA}. It is evident that the $P$-variable admits a generalization to the case of non equal masses.

\subsubsection{Towards $d=2$ and $d=1$}

For the operator ${\De_P}$ (\ref{OPDP}) the passage to lower dimensions is non-singular. Coming to $d=1,2$ the overall multiplicative factor in front of the first derivative term in (\ref{OPDP}) changes only.

As for the operator $\Delta_{q,P}$ in (\ref{DRP}), in the case $d=2$ it must involve four $q$-variables only
\[
\Delta_{q,P}|_{d=2} \ = \ \sum_{i+j+k+\ell=1}^{2} Y_{i,j,k,\ell}\,\partial^i_{q_1}\pa^j_{q_2}\partial^k_{q_3}\pa^\ell_{q_4} \ ,
\]
with certain coefficients $Y_{i,j,k,\ell}=Y_{i,j,k,\ell}(P,q_1,q_2,q_3,q_4)$. In general, $\Delta_{q,P}|_{d=2}$ is not an algebraic operator.

At $d=1$, the operator $\Delta_{q,P}$ depends on two $q$-variables alone
\[
\Delta_{q,P}|_{d=1} \ = \ \sum_{i+j=1}^{2} Z_{i,j}\,\partial^i_{q_1}\pa^j_{q_2} \ ,
\]
here $Z_{i,j}=Z_{i,j}(P,q_1,q_2)$. Again, in general $\Delta_{q,P}|_{d=1}$ is not algebraic.


\section{{(Quasi)}-exact-solvability}
\label{QESsec}

In this section, for $d\geq3$ we describe in more detail the exact and quasi-exactly solvable (QES) models for the four-body problem in the $\rho$-representation (space of relative distances).

\subsection{QES in $\rho$-variables, $d \geq 3$}

{\bf (I).\ \it Quasi-Exactly-Solvable problem in $\rho$-variables.}

Let us take the $d$-independent function
\begin{equation}
\label{psi_cal-r-d3}
       \Psi_0(\rho) \ \equiv \ F_2^{\frac{1}{4}}\,F_1^{\frac{\gamma}{2}} \,e^{-\om\,P - \frac{A}{2}\,P^2}\ ,
\end{equation}
where $\gamma,\,\om > 0$ and $A \geq 0$ and for $\om=0$, $A>0$ are constants. Here $P$ is given by (\ref{VarP}) and
\begin{equation}
\label{F1new}
F_1\ =\ {\cal V}\ ,
\end{equation}
\begin{equation}
\label{F2new}
   F_2\ =\ 36\,{\cal V}\ -\ P \, S\ ,
\end{equation}
are written in terms of the volume variables (\ref{V4})-(\ref{VarP}).
We look for the potential for which the function (\ref{psi_cal-r-d3}) is the ground state function
then for the Hamiltonian ${H}_{LB}(\rho)$ (\ref{H-3-3r-rho}) of the 6-dimensional quantum particle. This potential can be found immediately by calculating the ratio
\[
\frac{\De_{LB}(\rho) \Psi_0 }{ \Psi_0}\ =\ V_0 - E_0 \ ,
\]
where $\De_{LB}(\rho)$ is given by (\ref{LBop}) with metric (\ref{gmn33-rho}).
The  result is
\begin{equation}
V_0(\rho)\ = \ \frac{3\,P^2
+112\,S}{32\,F_2} \ + \ \gamma(\gamma-1)\frac{S}{18\,F_1}\ + \ 8\,\om^2\,P\, +\, 4\,A\,P\,(4\,\om\,P -\, 6\,\gamma - 11)\,
   +\, 8\, A^2\,P^3      \ ,
\label{VQES-0}
\end{equation}
which is $d$-independent, it includes both the effective potential $V_{eff}$ and many-body potential $V$ with the energy of the ground state
\begin{equation}
  E_0\ =\ 12\,\om\,(3+2\,\gamma) \ .
\label{EQES-0}
\end{equation}

Now, let us take the Hamiltonian ${H}_{LB,0} \equiv  -\De_{LB}(\rho) + V_0$ with potential (\ref{VQES-0}), subtract $E_0$ (\ref{EQES-0}) and make the gauge rotation with $\Psi_0$ (\ref{psi_cal-r-d3}). As the result we obtain the $sl(7, {\bf R})$-Lie-algebraic operator with additional potential $\De V_N$
\begin{equation}
\begin{aligned}
    & \Psi_0^{-1}\,(-{\De_{LB}}(\rho) + V_0 - E_0)\,\Psi_0   \  \equiv \ h^{(qes)}(J)\ +\ \De V_N\  = \ -{\De_R}({\cal J})
    \\ &
       +\ 2\,(d - 3 - 2\,\gamma)\,({\cal J}_1^- + {\cal J}_2^- + {\cal J}_3^-+ {\cal J}_4^-+ {\cal J}_5^-+ {\cal J}_6^-)
\\ &
     + \ 16\,A\,\left({\cal J}_1^+(N) + {\cal J}_2^+(N)  + {\cal J}_3^+(N)+ {\cal J}_4^+(N)+ {\cal J}_5^+(N)+ {\cal J}_6^+(N) \right)
\\ &   + \ 16\,\om\,({\cal J}_{11}^0 \, + \, {\cal J}_{22}^0 \, + \, {\cal J}_{33}^0\, +\,  {\cal J}_{44}^0\, + \, {\cal J}_{55}^0\,  + \, {\cal J}_{66}^0) \ + \ \De V_N \ ,
\label{HQES-0-Lie}
\end{aligned}
\end{equation}
see (\ref{HRex}), where
\[
    \De V_N\ = \ 16 \,A\,N\,P \ =\ 16 \,A\,N\,(\rho_{12} \,+\, \rho_{13} \,+\, \rho_{14} \,+\, \rho_{23} \,+\, \rho_{24} \,+\, \rho_{34} \, )\ .
\]
It is evident that if the parameter $N$ takes integer value, the $d$-independent operator $h^{(qes)}(J)$ has a finite-dimensional invariant subspace ${\cal P}_{N}$, (\ref{P3})
with $\dim {\cal P}_{N} \sim N^3$ at large $N$.
Finally, we arrive at the quasi-exactly-solvable, $d$-independent, single particle Hamiltonian in the space of relative distances $\rho$,
\begin{equation}
\label{HQES-3-3}
    {H}_{LB}^{(qes)}(\rho) \ =\ -\De_{LB}(\rho) \ + \  V_N^{(qes)}(\rho)\ ,
\end{equation}
cf.(\ref{Hrel-final}), where
\begin{equation}
   V^{(qes)}_N =  \frac{3\,P^2
+112\,S}{32\,F_2} \, + \, \frac{\gamma(\gamma-1)S}{18\,F_1}\, + \, 8\,\om^2\,P\, +\, 4\,A\,P\,(4\,\om\,P - 6\,\gamma  -  11  - 4\,N)\,
   +\, 8\, A^2\,P^3      \ ,
\label{VQES-N}
\end{equation}
is a QES potential. Its configuration space is defined by $F_1 \geq 0$, while if it is fulfilled then $F_2\geq 0$.

For this potential $\sim N^3$ eigenstates can be found by algebraic means. They have the factorized form of the polynomial in $\rho$ multiplied by $\Psi_0$ (\ref{psi_cal-r-d3}),
\begin{equation}\label{Polh}
  \mbox{Pol}_N (\rho_{12}, \rho_{13},\rho_{14},\rho_{23},\rho_{24},\rho_{34})\ \Psi_0 (F_1,\, F_2,\,P)
  \ .
\end{equation}
These polynomials are the eigenpolynomials of the quasi-exactly-solvable, $d$-independent, algebraic operator
\begin{equation}
\begin{aligned}
\label{hQES-N-alg}
& \frac{1}{2}\,h^{(qes)}(\rho)\ =   \  - 2\,(\rho_{12} \,\pa^2_{\rho_{12}} \,+\, \rho_{13}\, \pa^2_{\rho_{13}}\, +\,\rho_{14}\, \pa^2_{\rho_{14}}\,+\,\rho_{23}\, \pa^2_{\rho_{23}} \,+\,\rho_{24}\, \pa^2_{\rho_{24}} \,+\,\rho_{34}\, \pa^2_{\rho_{34}} )
\\ &
- \ \big((\rho_{12} + \rho_{13} - \rho_{23})\pa_{\rho_{12}}\pa_{\rho_{13}}\ +
          (\rho_{12} + \rho_{14} - \rho_{24})\pa_{\rho_{12}}\pa_{\rho_{14}}\ +
          (\rho_{13} + \rho_{14} - \rho_{34})\pa_{\rho_{13}}\pa_{\rho_{14}} \big)
\\ &
- \  \big((\rho_{12} + \rho_{23} - \rho_{13})\pa_{\rho_{12}}\pa_{\rho_{23}}\ +\
          (\rho_{12} + \rho_{24} - \rho_{14})\pa_{\rho_{12}}\pa_{\rho_{24}}\ +\
          (\rho_{23} + \rho_{24} - \rho_{34})\pa_{\rho_{23}}\pa_{\rho_{24}}
    \big)
\\ &
-  \ \big((\rho_{13} + \rho_{23} - \rho_{12})\pa_{\rho_{13}}\pa_{\rho_{23}}\ +\
          (\rho_{13} + \rho_{34} - \rho_{14})\pa_{\rho_{13}}\pa_{\rho_{34}}\ +\
          (\rho_{23} + \rho_{34} - \rho_{24})\pa_{\rho_{23}}\pa_{\rho_{34}}
    \big)
\\ &
- \  \big((\rho_{14} + \rho_{24} - \rho_{12})\pa_{\rho_{14}}\pa_{\rho_{24}}\ +\
          (\rho_{14} + \rho_{34} - \rho_{13})\pa_{\rho_{14}}\pa_{\rho_{34}}\ + \
          (\rho_{24} + \rho_{34} - \rho_{23})\pa_{\rho_{24}}\pa_{\rho_{34}}
    \big)
\\ &
- \ (2\,\gamma \,+\, 3)\,(\pa_{\rho_{12}} + \pa_{\rho_{13}}+ \pa_{\rho_{14}}+ \pa_{\rho_{23}}+ \pa_{\rho_{24}}+ \pa_{\rho_{34}})
\\ &
 + \  8\,\omega\,(\rho_{12}\,\pa_{\rho_{12}} + \rho_{13}\,\pa_{\rho_{13}}+ \rho_{14}\,\pa_{\rho_{14}}+ \rho_{23}\,\pa_{\rho_{23}}+ \rho_{24}\,\pa_{\rho_{24}}+ \rho_{34}\,\pa_{\rho_{34}})
\\ &
 + \  8\,A\,P\,(\rho_{12}\,\pa_{\rho_{12}} + \rho_{13}\,\pa_{\rho_{13}}+ \rho_{14}\,\pa_{\rho_{14}}+ \rho_{23}\,\pa_{\rho_{23}}+ \rho_{24}\,\pa_{\rho_{24}}+ \rho_{34}\,\pa_{\rho_{34}} \ - \ N) \ ,
\end{aligned}
\end{equation}
or, equivalently, of the quasi-exactly-solvable $sl(7,\,{\bf R})$-Lie-algebraic operator
\begin{equation}
\label{hQES-N-Lie}
\begin{aligned}
 \frac{1}{2}\,h^{(qes)}(J) & \ = \ -2\,(\, {\cal J}_{11}^0\,{\cal J}_1^- + {\cal J}_{22}^0\,{\cal J}_2^-
      + {\cal J}_{33}^0\,{\cal J}_3^-  + {\cal J}_{44}^0\,{\cal J}_4^-  + {\cal J}_{55}^0\,{\cal J}_5^-  + {\cal J}_{66}^0\,{\cal J}_6^-  \,)\
\\ &
-  \bigg[{\cal J}_{11}^0\,({\cal J}_2^- + {\cal J}_3^-  + {\cal J}_4^-  + {\cal J}_5^- ) + {\cal J}_{22}^0\,({\cal J}_1^- + {\cal J}_3^-  + {\cal J}_4^-  + {\cal J}_6^- )
\\ &
 +  {\cal J}_{33}^0\,({\cal J}_1^- + {\cal J}_2^-  + {\cal J}_5^-  + {\cal J}_6^- )
+ {\cal J}_{44}^0\,({\cal J}_1^- + {\cal J}_2^-  + {\cal J}_5^-  + {\cal J}_6^- )
\\ &
+ {\cal J}_{55}^0\,({\cal J}_1^- + {\cal J}_3^-  + {\cal J}_4^-  + {\cal J}_6^- ) +  {\cal J}_{66}^0\,({\cal J}_2^- + {\cal J}_3^-  + {\cal J}_4^-  + {\cal J}_5^- ) \bigg]
\\ &
+\ 2\, \bigg[ {\cal J}_{12}^0\,{\cal J}_4^-  +  {\cal J}_{13}^0\,{\cal J}_5^-   +  {\cal J}_{21}^0\,{\cal J}_4^-   +   {\cal J}_{23}^0\,{\cal J}_6^-   +  {\cal J}_{31}^0\,{\cal J}_5^-   +   {\cal J}_{32}^0\,{\cal J}_6^-
\\ &
+   {\cal J}_{41}^0\,{\cal J}_2^-   +  {\cal J}_{45}^0\,{\cal J}_6^- + {\cal J}_{54}^0\,{\cal J}_6^-  +  {\cal J}_{62}^0\,{\cal J}_3^-  +   {\cal J}_{64}^0\,{\cal J}_5^-  +   {\cal J}_{51}^0\,{\cal J}_3^-        \bigg]
\\ &
\ -\ (3\, +2\,\gamma)\,( {\cal J}_1^- + {\cal J}_2^- + {\cal J}_3^- + {\cal J}_4^- + {\cal J}_5^- + {\cal J}_6^- )
\\  &
+ \ 8\,A\,\left({\cal J}_1^+(N) + {\cal J}_2^+(N)  + {\cal J}_3^+(N)+ {\cal J}_4^+(N)+ {\cal J}_5^+(N)+ {\cal J}_6^+(N) \right)
\\ &
   + \ 8\,\om\,({\cal J}_{11}^0 \, + \, {\cal J}_{22}^0 \, + \, {\cal J}_{33}^0\, +\,  {\cal J}_{44}^0\, + \, {\cal J}_{55}^0\,  + \, {\cal J}_{66}^0)  \ ,
\end{aligned}
\end{equation}
cf. (\ref{HQES-0-Lie}).

As for the original many-body problem (\ref{Hrel-Mod}) in the space of relative distances
\[
   {{\cal H}}_{r}\,\Psi(r) \equiv \ \bigg(- {\De_{{radial}}}(r)\ + \ V(r)\bigg)\, \Psi(r)\ =\ E\,\Psi(r)\ ,\quad \Psi \in L_2 ({\Re_{{radial}}})\ ,
\]
the potential for which quasi-exactly-solvable, polynomial solutions occur in the form
\[
   \mbox{Pol}_N (\rho_{12}, \rho_{13},\rho_{14},\rho_{23},\rho_{24},\rho_{34})\ \Gamma(F_1,\,F_2)\ \Psi_0 (F_1,\, F_2,\,P)\ ,
\]
where $\Gamma \sim   D^{-1/4}\,F_1^{\frac{4-d}{4}}$, see (\ref{GamFac}) is given by
\[
V_{relative, N}^{(qes)} \ = \   V^{(qes)}_N \ - \ V_{eff}\ =
\]

\begin{equation}
\frac{4\,\gamma(\gamma-1)-(d-5)(d-3)}{72}\frac{S}{F_1}  \ + \ 8\,\om^2\,P\, +\, 4\,A\,P\,(4\,\om\,P -\, 6\,\gamma \, - \, 11 \, -\, 4\,N)\,
   +\, 8\, A^2\,P^3      \ ,
\label{VQES-N-rel}
\end{equation}
c.f. (\ref{VQES-N}); it does not depend on $F_2$ and does not contain a singular term $\sim 1/F_2$.

\bigskip

{\bf (II).\ \it Exactly-Solvable problem in $\rho$-variables.}

If the parameter $A$ vanishes in (\ref{psi_cal-r-d3}), (\ref{VQES-N}) and (\ref{HQES-0-Lie}), (\ref{hQES-N-Lie}) we will arrive at the
exactly-solvable problem, where $\Psi_0$ (\ref{psi_cal-r-d3}) at $A=0$ plays the role of the ground state function,
\begin{equation}
\label{psi_cal-r-d2exact}
   \Psi_0(\rho_{12},\,\rho _{13},\,\rho _{23}) \ = \ F_2^{\frac{1}{4}}\,F_1^{\frac{\gamma}{2}} \,e^{-\om\,P } \ .
\end{equation}
The $sl(7, {\bf R})$-Lie-algebraic operator (\ref{hQES-N-Lie}) contains no raising generators $\{{\cal J}^+(N)\}$ and becomes
\begin{equation}\label{}
\begin{aligned}
h^{(exact)} =& -{\De_R}({\cal J})\
   +\ 2\,(d - 3 - 2\,\gamma)\,({\cal J}_1^- + {\cal J}_2^- + {\cal J}_3^-+ {\cal J}_4^-+ {\cal J}_5^-+ {\cal J}_6^-)
   \\ &
+ \ 16\,\om\,({\cal J}_{11}^0 \, + \, {\cal J}_{22}^0 \, + \, {\cal J}_{33}^0\, +\,  {\cal J}_{44}^0\, + \, {\cal J}_{55}^0\,  + \, {\cal J}_{66}^0) \ ,
\end{aligned}
\end{equation}
see (\ref{HRex}), and, hence, preserves the infinite flag of finite-dimensional invariant subspaces
${\cal P}_{N}$ (\ref{P3}) at $N=0,1,2\ldots$\,. The single particle potential (\ref{VQES-N}) becomes

\begin{equation}
 V^{(es)}(\rho)\ = \ \frac{3\,P^2
+112\,S}{32\,F_2} \ + \ \gamma(\gamma-1)\frac{S}{18\,F_1} \  + \ 8\,\om^2\,P \ .
\label{VES}
\end{equation}
Eventually, we arrive at the exactly-solvable single particle Hamiltonian in the space of relative distances
\begin{equation}
\label{HES-3-2}
    {H}_{LB}^{(es)}(\rho) \ =\ -\De_{LB}(\rho) \ + \  V^{(es)}(\rho)\ ,
\end{equation}
where the spectra of energies
\[
    E_{N}\ =\ 12\,\om\,(N+3+2\,\gamma)\ ,\qquad N=0,1,2,\ldots \qquad ,
\]
is equidistant. Its degeneracy is equal to the number of partitions of $$N=n_1 + n_2 + n_3 +n_4+n_5+n_6\ .$$
All eigenfunctions have the factorized form of a polynomial in $\rho$ multiplied by $\Psi_0$ (\ref{psi_cal-r-d2exact}),
\[
  \mbox{Pol}_N (\rho_{12}, \rho_{13},\rho_{14},\rho_{23},\rho_{24},\rho_{34})\ \Psi_0 (F_1,\, F_2,\,P)  \ ,\qquad N=0,1,2,\ldots\ .
\]

Note these polynomials are eigenpolynomials of the exactly-solvable, $d$-independent, algebraic operator (\ref{hQES-N-Lie}) with $A=0$,
\[
    h^{(exact)}(\rho)\ = \ h^{(qes)}(\rho)\mid_{A=0} \ .
\]
The polynomials $\mbox{Pol}_N$ are orthogonal w.r.t. $\Psi_0^2$ (\ref{psi_cal-r-d2exact}) in domain  given by (\ref{CFrho}). To the best of our knowledge these orthogonal polynomials have not been studied in literature.

The Hamiltonian with potential (\ref{VES}) can be considered as a type of a $d$-dimensional generalization of the 4-body Calogero model \cite{Calogero:1969} with loss of the property of pairwise interaction only. Now the potential of interaction contains two-, three- and four-body interactions. If $\gamma=0,1$ in (\ref{VES}) we arrive at the celebrated harmonic oscillator potential in the space of relative distances, see e.g. \cite{Green}-\cite{Moshinsky} for the three-body case. In turn, in the space of relative motion this potential contains no singular terms at all and becomes,
\begin{equation}
\label{harmonic}
     V_{harmonic} \ =\ 8\,\om^2\,P\ =\ 8\,\om^2\,(\rho_{12} + \rho_{13} + \rho_{14}+\rho_{23} + \rho_{24} + \rho_{34})\ .
\end{equation}
We arrive at the (non-singular) harmonic oscillator potential $V_{harmonic}$. The potential (\ref{VES}) is a $d$-dimensional generalization of the harmonic oscillator in the space of relative motion
rather than a potential of generalized four-body (rational) Calogero model.


\section*{Conclusions}

In this paper we studied the quantum four body problem in a $d$-dimensional space. Based on the change of variables from individual Cartesian coordinates $\{ {\bf r}_i \}$ to centre-of-mass vector coordinate ${\bf R}_{CM}$, mutual relative distances between bodies $\{ r_{ij} \}$ and angles $\{ {\Omega} \}$,
\[
    (\,{\bf r}_1,\,{\bf r}_2,\,{\bf r}_3,\,{\bf r}_4\,) \quad  \Leftrightarrow  \quad \big(\,{\bf R}_{CM},\, \{ r_{ij} \},\, \{ \Omega \}\, \big)\ ,
\]
the kinetic energy given by the original flat diagonal Laplace operator decomposes naturally into the sum of three operators
\[
 \sum_{i=1}^4\frac{1}{2} \Delta_i^{(d)}\ =\ \Delta_{{\bf R}_{CM}} \ + \  \Delta_{radial} \ + \ \Delta_{\Omega} \ ,
\]
where $\Delta_{{\bf R}_{CM}}$ is the center of mass Laplacian, the operator $\Delta_{radial}$
depends on the mutual distances (equivalently, the radial variables) only, ${\rho_{ij}=r_{ij}^2}$, and $\De_{\Om}$ annihilates any function of the radial variables alone. The
operator $\Delta_{radial}(\rho)$ is self-adjoint, it does not depend on how angular variables $\Om$ are introduced. It is positive-definite. Also it is an $sl(7, R)$-Lie-algebraic operator, see (\ref{addition3-3rho}) and (\ref{HRex}).

On the subspace of the Hilbert space of angle-independent eigenfunctions, the above-mentioned change of variables implies that the original multi-dimensional spectral problem,
\[
    {\cal H}\, \Psi \ =\ E\, \Psi \ ,
\]
is reduced to a much simpler, {\it restricted} one,
\begin{equation}
\label{restricted}
   \bigg(\,-\Delta_{radial}(\rho) \, +\, V(\rho)\,\bigg)\, \psi \ =\ E\, \psi \ .
\end{equation}
This restricted spectral problem depends on six variables solely. Moreover, the ground state function, if it exists, should be an eigenfunction of such restricted spectral problem as was predicted by Ter-Martirosyan \cite{Ter}.

It was shown that there exists a gauge factor $\Gamma$ such that the l.h.s. in (\ref{restricted}) is gauge-equivalent to the Hamiltonian
of a six-dimensional quantum particle in a curved space in external potential,
\[
 {\cal H}_{LB}  \ \equiv \ \Gamma^{-1} \bigg( -\Delta_{radial}(\rho)\, +\, V(\rho)\bigg) \Gamma \ =\ -\Delta_{LB}\, +\, V_{eff}(\rho)\, +\, V(\rho)
   \ .
\]
Here $\Delta_{LB}$ is the Laplace-Beltrami operator with contravariant metric $g^{\mu \nu}$ (\ref{gmn33-rho}), and $V_{eff}(\rho)$ (\ref{VeffF1}) is the effective potential which emerged as a result of $\Gamma$-gauge rotation . The boundary of the configuration space for ${\cal H}_{LB}$ is defined by the condition $\det g^{\mu \nu}=0$.

The (Lie)-algebraic form of the operator $\Delta_{radial}(\rho)$ suggests way to finding the exact solutions of both restricted and original spectral problems. In particular, adding to $\Delta_{radial}(\rho)$ the terms linear in derivatives, $A_{ij}\, \rho_{ij}\, \pa_{ij}$, and then gauging them away with factor $\sim \exp ( - {\tilde A}_{ij} \,\rho_{ij})$ leads to the anisotropic harmonic oscillator potential in the space of relative distances,
\[
     V^{(ex)}\ =\ \sum_{i<j}^{6}\om^2_{ij} \, \rho_{ij}\ ,
\]
which is an exactly-solvable potential for the restricted problem and perhaps, quasi-exactly-solvable for the original problem.

A novel result was the introduction of two different representations for the operator $\Delta_{radial}$ in (\ref{restricted}). They involve pure geometrical variables defined by the tetrahedron of interaction. In particular, the volume-variables representation allows us a better understanding of the degeneration from $d\geq3$ to lower dimensions $d=2$ and $d=1$. In this limiting process, a Lie-algebraic sector of the problem is preserved. For the restricted problem (\ref{restricted}) in the volume-variables representation we arrive, provided that the original potential only depends on the volume variables, at the gauge-equivalent Hamiltonian
\begin{equation}
    {\cal H}_{LB} ({\cal V},S,P) \ =\ -\De_{LB}({\cal V},S,P)\  + \  \tilde V_g({\cal V},S,P) \  + \   V({\cal V},S,P)   \ ,
\end{equation}
which describes a three-dimensional quantum particle moving in a curved space.

Interestingly, in the $u$-variables representation there exists another gauge-equivalent Hamiltonian
\begin{equation}
\label{HLBCC}
    {\cal H}_{LB} (u_1,u_2,u_3) \ =\ -\De_{LB}(u_1,u_2,u_3)\  + \  \tilde V_u(u_1,u_2,u_3) \  + \   V(u_1,u_2,u_3)   \ ,
\end{equation}
in the space of $u$-variables which describes a three-dimensional quantum particle moving not in a curved but in a flat space. At $d=1$ the operator (\ref{HLBCC}), after a suitable gauge rotation and change of variables, reduces to the four-body ($A_3$) rational Calogero-Sutherland model.

For any $d$, in the $P$-variable representation we have the remarkable property of the existence of a family of eigenfunctions of the four-body problem that only depend on the $P$-variable.

Consequently, exactly- and quasi-exactly-solvable models can be constructed for any $d$. This reveals interesting links between exact solvability and polyhedra which, more importantly, set up the basis towards the \emph{geometrization} of the $n$-body problem. The question about the existence of a representation in which the whole operator $\Delta_{radial}$ in (\ref{restricted}) remains algebraic at $d=2$ is still open.

Also, the case of non-equal masses is presented in the Appendix A. The operator $\Delta_{radial}\rightarrow \Delta^\prime_{radial}$ (\ref{addition3-3r-M}) admits a simple limit to the atomic (say, $m_1 \to \infty$) and molecular
(say, $m_{1,\ldots,p} \to \infty$) situations. In the atomic case, for the operator $\De^\prime_{radial}(\rho)$ (\ref{addition3-3r-M}) all second order cross derivatives $\pa_{\rho_{1j}}\pa_{\rho_{1k}}$ disappear, while other terms remain. The number of variables in this case remains unchanged. In the molecular case, not only cross derivatives $\pa_{\rho_{qj}}\pa_{\rho_{qk}}, q=1,\ldots,p$ but also the derivatives w.r.t. $\rho_{ij},\ 1 \leq i<j \leq p$ vanish. Thus, in general the operator $\De^\prime_{radial}$ depends on $6 - \frac{p(p-1)}{2}$ variables. Other variables which may appear in the potential $V(\rho)$ are external parameters. It corresponds to the so-called Bohr-Oppenheimer approximation (of zero order) in molecular physics.

In the Appendix B, we introduce the volume variables for the case of arbitrary masses. In the Appendix C, the generalization of the volume-variables to the $n$-body case is presented as well.

\section*{Acknowledgments}

A.V.T. is thankful to University of Minnesota, USA for kind hospitality extended to him where this work was initiated and to CRM, Montreal, Canada where it was continued during its numerous visits.
W.M. was partially supported by a grant from the Simons Foundation (\# 412351 to Willard Miller, Jr.).
The research of  M.A.E.R. was partially supported by a fellowship awarded by the Laboratory of Mathematical Physics of the CRM.  M.A.E.R. is grateful to ICN UNAM, Mexico for the kind hospitality during his visits, where the work was continued and eventually completed.

\appendix

\section{$\rho$-representation for non-equal masses}

Consider the general case of four particles located at points ${\bf r}_1,{\bf r}_2,{\bf r}_3,{\bf r}_4$
of masses $m_1,m_2,m_3,m_4$, respectively. The analogue of decomposition of kinetic energy of relative motion $\De_{rel}^{(3d)}$, see (\ref{Hgen}),(\ref{Hrel}), in variables $(r_{ij}, \Om)$ exists,
\[
      \frac{1}{2}\De_{rel}^{(3d,m)}\ =\ {\De_{radial}^{(6,m)}}(r_{ij}, \pa_{ij}) \ + \
    {\De}_{\Om}^{(3d-6,m)} (r_{ij}, \Om, \pa_{ij}, \pa_{\Om})
    \ ,\qquad \pa_{ij} \equiv \frac{\pa}{\pa_{r_{ij}}}\ ,
\]
cf.(\ref{addition}). Explicitly, the operator $\De_{radial}^{(6,m)}(r_{ij}, \pa_{ij})$ becomes (in terms of the relative coordinates $\rho_{ij}=r_{ij}^2$), see \cite{MTE:2018},
{\small
\begin{equation}
\begin{aligned}
\label{addition3-3r-M}
& { \De^\prime_{radial}}(\rho_{ij}, \pa_{ij})\ =  \   2\bigg( \frac{1}{\mu_{12}} \rho_{12} \,\pa^2_{\rho_{12}} + \frac{1}{\mu_{13}}\rho_{13}\, \pa^2_{\rho_{13}} +\frac{1}{\mu_{14}}\rho_{14}\, \pa^2_{\rho_{14}}
 +\frac{1}{\mu_{23}}\rho_{23}\, \pa^2_{\rho_{23}} +\frac{1}{\mu_{24}}\rho_{24}\, \pa^2_{\rho_{24}}
 \\ &  +\frac{1}{\mu_{34}}\rho_{34}\, \pa^2_{\rho_{34}} \bigg) + d\,\bigg(\frac{1}{\mu_{12}}\pa_{\rho_{12}} + \frac{1}{\mu_{13}}\pa_{\rho_{13}}+ \frac{1}{\mu_{14}}\pa_{\rho_{14}}+ \frac{1}{\mu_{23}}\pa_{\rho_{23}}+ \frac{1}{\mu_{24}}\pa_{\rho_{24}}+ \frac{1}{\mu_{34}}\pa_{\rho_{34}}\bigg)
\\ &
 +  \frac{2}{m_1} \bigg({(\rho_{12} + \rho_{13} - \rho_{23})}\pa_{\rho_{12}}\pa_{\rho_{13}}\ +
          {(\rho_{12} + \rho_{14} - \rho_{24})}\pa_{\rho_{12}}\pa_{\rho_{14}}\ +
          {(\rho_{13} + \rho_{14} - \rho_{34})}\pa_{\rho_{13}}\pa_{\rho_{14}} \bigg)
\\ &
+  \frac{2}{m_2} \bigg((\rho_{12} + \rho_{23} - \rho_{13})\pa_{\rho_{12}}\pa_{\rho_{23}}\ +
          (\rho_{12} + \rho_{24} - \rho_{14})\pa_{\rho_{12}}\pa_{\rho_{24}}\ +
          (\rho_{23} + \rho_{24} - \rho_{34})\pa_{\rho_{23}}\pa_{\rho_{24}}
    \bigg)
\\ &
+  \frac{2}{m_3} \bigg((\rho_{13} + \rho_{23} - \rho_{12})\pa_{\rho_{13}}\pa_{\rho_{23}}\ +
          (\rho_{13} + \rho_{34} - \rho_{14})\pa_{\rho_{13}}\pa_{\rho_{34}}\ +
          (\rho_{23} + \rho_{34} - \rho_{24})\pa_{\rho_{23}}\pa_{\rho_{34}}
    \bigg)
\\ &
+  \frac{2}{m_4} \bigg((\rho_{14} + \rho_{24} - \rho_{12})\pa_{\rho_{14}}\pa_{\rho_{24}}\ +
          (\rho_{14} + \rho_{34} - \rho_{13})\pa_{\rho_{14}}\pa_{\rho_{34}}\ +
          (\rho_{24} + \rho_{34} - \rho_{23})\pa_{\rho_{24}}\pa_{\rho_{34}}
    \bigg)      \ ,
\end{aligned}
\end{equation}
}
cf.(\ref{addition3-3rho}) for the case of equal masses $m_1=m_2=m_3=m_4=1$, where

\[
   \frac{1}{\mu_{ij}}\ =\ \frac{m_i+m_j}{m_i m_j}\ ,
\]
is the reduced mass for particles $i$ and $j$.
This operator has the same algebraic structure as ${\De_{radial}}(\rho_{ij})$
but lives on a different manifold in general. It can be rewritten in terms of the
generators of the maximal affine subalgebra $b_7$ of the algebra $sl(7,{\bf R})$,
see (\ref{sl4R}), c.f. (\ref{HRex}). The contravariant metric tensor, obtained from the coefficients in front of the second derivatives in (\ref{addition3-3r-M}), does not depends on $d$ and
its determinant is

\begin{equation}
\label{gmn33-rho-det-M}
D_m\ =\ \det g^{\mu \nu}\ =\ 9216\,c_m\,V_4^2\,\bigg[ \big(\sum V_{2,m}\big)\big( \sum V_{3,m}\big)\, -\, 9\,(m_1+m_2+m_3+m_4)\,V_4^2      \bigg]  \ ,
\end{equation}
and is positive definite, where $c_m=\frac{m_1+m_2+m_3+m_4}{m_1^2\,m_2^2\,m_3^2\,m_4^2}$, $V_4^2$ given by (\ref{V4}) and
\[
\sum V_{2,m}\ =\ m_1 m_2 r_{12}^2+m_1 m_3 r_{13}^2+m_1 m_4 r_{14}^2+m_2 m_3 r_{23}^2+m_2 m_4 r_{24}^2+m_3 m_4 r_{34}^2 \ ,
\]
is the weighted sum of square of sides and diagonals of the tetrahedron of interaction and
{\small
\[
\sum V_{3,m} \ = \  \frac{1}{m_1} S^2(r_{23},\,r_{24},\,r_{34}) \ + \ \frac{1}{m_2} S^2(r_{13},\,r_{14},\,r_{34})\ + \  \frac{1}{m_3} S^2(r_{12},\,r_{14},\,r_{24})\ + \ \frac{1}{m_4} S^2(r_{12},\,r_{13},\,r_{23}) \ ,
\]
}
is the weighted sum of squares of areas, $ S^2(a,\,b,\,c)$ is the square of the area of the triangle of interaction with sizes $a,b,c$. Hence, $D_m$ is still proportional to the squared of the volume of tetrahedron $V_{4}^2$ being of pure geometrical nature!

Making the gauge transformation of (\ref{addition3-3r-M}) with determinant (\ref{gmn33-rho-det-M}) as the factor,
\[
         \Gamma \ = \  D_m^{-\frac{1}{4}}\,V_4^{1 - \frac{d}{4}} \quad ,
\]
we find that
\begin{equation}
         \Gamma^{-1}\, {\De'_{radial}}(\rho_{ij})\,\Gamma \ =
        \  \De'_{LB}(\rho) \ - \ V_{eff} \ ,
\label{HLB3Ma}
\end{equation}
is the Laplace-Beltrami operator plus the effective potential
\[
V_{eff}\ =\  \ \frac{3\,(\sum {V}_{2,m}^2)^2+28\,(m_1+m_2+m_3+m_4)\,m_1\,m_2\,m_3\,m_4\,\sum{ V}_{3,m}^2}{32\,m_1\,m_2\,m_3\,m_4\,((\sum { V}_{2,m}^2)\sum { V}_{3,m}^2-9\,(m_1+m_2+m_3+m_4)\,{ V}_4^2)}
\]
\begin{equation}
+ \ \frac{(d-5)(d-3)\sum{V}_{3,m}^2}{72\,{V}_4^2}   \ ,
\label{Veffm}
\end{equation}
where its second term is absent for $d=3,5$.
The Laplace-Beltrami operator plays a role of the kinetic energy of six-dimensional quantum particle moving in curved space. While $V_{eff}$ reminds the centrifugal potential.

\section{volume-variables representation for non-equal masses}

For arbitrary masses $(m_1\,, m_2\,, m_3\,, m_4)$\,, the analogue of decomposition (\ref{DEsum})
for modified by arbitrary masses $\Delta^\prime_{radial}$ can be written and the analogue of the operator ${\De}_g$ (\ref{DEg}) can be derived in modified volume variables,
\begin{equation}
\begin{aligned}
{\label{DEgmasses}}
   {\De}^\prime_g  \ = & \   \frac{2}{9}\,{\cal V}\,\tilde S\,\pa^2_{{\cal V},{\cal V}}  \ +  \
   \big(  \frac{27\,M}{2 \,m}\,{\cal V} \, + \, \frac{1}{2\,m}\tilde S\,\tilde P \  \big)\,\pa^2_{\tilde S,\tilde S} \ + \ 2\,M\,\tilde P\,\pa^2_{\tilde P,\tilde P}
\\ &  +  \  8\,M\,\,\tilde S\,\pa^2_{\tilde S,\tilde P} \  +
   \ 2\,{\cal V}\,\big(    \frac{1}{m}\tilde P\,\pa^2_{{\cal V},\tilde S} \ +
   \ 6\,M\,\,\pa^2_{{\cal V},\tilde P} \  \big) \ + \ \frac{1}{9}(d-2)\,\tilde S\,\pa_{{\cal V}}
\\ &  + \
   \frac{1}{2\,m}(d-1)\,\tilde P\,\pa_{\tilde S} \ + \ 3\,M\,d\,\pa_{\tilde P}\ ,
\end{aligned}
\end{equation}
where $M=m_1+m_2+m_3+m_4$, $m=m_1\,m_2\,m_3\,m_4$\ , and

\[
 {\cal V} \ \equiv \ V_4^2 \ ,
\]
\[
\tilde P  \equiv  \sum V_{2,m}\ =\ m_1 m_2 r_{12}^2+m_1 m_3 r_{13}^2+m_1 m_4 r_{14}^2+m_2 m_3 r_{23}^2+m_2 m_4 r_{24}^2+m_3 m_4 r_{34}^2 \ ,
\]
\[
\tilde S \ \equiv \ \sum V_{3,m} \ = \  \frac{1}{m_1} S^2(r_{23},\,r_{24},\,r_{34}) \ + \ \frac{1}{m_2} S^2(r_{13},\,r_{14},\,r_{34})
\]
\begin{equation}
\label{volume-v-m}
   + \  \frac{1}{m_3} S^2(r_{12},\,r_{14},\,r_{24})\ + \ \frac{1}{m_4} S^2(r_{12},\,r_{13},\,r_{23})
   \ .
\end{equation}

The contravariant metric tensor obtained from (\ref{DEgmasses}) does not depends on $d$ and its determinant is

\begin{equation}
\label{DEgmassesdet}
D_{gm}\ =\ 2\,M\,{\cal V}\, \frac{ m\, \left(162 \,M \,\tilde P \,\tilde S \,{\cal V}-2187\, M^2\, {\cal V}^2+\tilde P^2\, \tilde S^2\right)-16\, m^2\, M\, \tilde S^3-9 \, \tilde P^3 \,{\cal V}}{9\, m^2}    \bigg]  \ .
\end{equation}

Making the gauge transformation of (\ref{DEgmasses}) with determinant (\ref{DEgmassesdet}) and volume of tetrahedron as the factor:
\[
         \Gamma \ = \  D_{gm}^{-1/4}\,{\cal V}^{1-\frac{d}{4}} \ ,
\]
we find that
\begin{equation}
         \Gamma^{-1}\, {{\De}^\prime_g }(\tilde P,\,\tilde S,\,{\cal V})\,\Gamma \ =
        \  \De'_{g,LB}(\tilde P,\,\tilde S,\,{\cal V}) \ - \ V_{eff} \ ,
\label{HLB3Mb}
\end{equation}
is the Laplace-Beltrami operator with the effective potential
\[
V_{eff}\ =\ \ \ \frac{\left(\tilde P^2-12\, m\, M\, \tilde S\right) (81 \,M\, {\cal V}-\tilde P \,\tilde S) }{8 \left(2187\, m\, M^2 {\cal V}^2+m\, \tilde S^2 \left(16\, m\, M \,\tilde S-\tilde P^2\right)+9 \,\tilde P\, {\cal V} \left(\tilde P^2-18 \,m \,M \,\tilde S\right)\right)}
\]
\begin{equation}
\label{Veffmb}
    + \ (d-5) (d-3)\frac{ \tilde S}{72 \,{\cal V}}  \ ,
\end{equation}
where the second term is absent for $d=3,5$.
The Laplace-Beltrami operator plays a role of the kinetic energy of three-dimensional quantum particle moving in curved space.

\section{Geometrical variables for the $n$-body system}

Based on concrete results for $n=2,3,4,5$ we introduce geometrical variables for the $n$-body system in $d$-dimensional space $d \geq n-1$. They allow to study the degeneration of the system from $d\geq n-1$ to lower dimensions.

\subsection{volume-variables representation for the $n$-body system}

For equal masses $m_i=1$ ($i=1,2,\ldots,n$), let us introduce the set of $(n-1)$ volume variables $\{{\cal V}_k\}$, $k=2,3,\ldots,n$, where ${\cal V}_{n}$ is the volume (squared) of the $n$-vertex polytope of interaction (whose vertices correspond to the positions of the particles) and ${\cal V}_k$ is the sum over the squares of the {contents} (volumes of faces) of fixed dimension $k$. In these variables, the operator $\Delta_{n,radial}$ \cite{MTE:2018} which depends solely on the $\frac{n(n-1)}{2}$ relative distances between particles can decomposed as the sum of two operators
\begin{equation}
\label{}
\Delta_{n,radial} \ = \ \Delta_{n,g} \ + \   \Delta_{n,q} \ ,
\end{equation}
($[\Delta_{n,g},\,\Delta_{n,q}] \neq 0 $) with the following properties
\begin{itemize}
  \item $\Delta_{n,g}=\Delta_{n,g}(\{{\cal V}_k\})$: it is an algebraic operator for any $d$, it involves {volume variables} $\{{\cal V}_k\}$, $k=2,3,\ldots,n$, alone. Explicitly,
\begin{equation}
\begin{aligned}
{\label{DVn}}
{\De}_{n,g}  \ = & \  {\cal V}_n \sum_{i=2}^{n-1} a_i\,{\cal V}_i\,\pa^2_{i+1,n}
 \    + \ \sum_{i=2}^{n} b_i\,{\cal V}_i\,\pa^2_{i,2} \  + \  \sum_{i=0}^{n-2}\,e_i\,(d-i)\,{\cal V}_{i+1}\,\pa_{i+2}
\\ & + \ \sum_{j=1}^{n-3}\sum_{i=1}^{j}\bigg( c_{i,j}\,{\cal V}_{n+1-i}\,{\cal V}_{n-j-2} \,+\,  f_{i,j}\,{\cal V}_{n-i}\,{\cal V}_{n-j-1}    \bigg)\,\pa^2_{n-i,n-j}   \ .
\end{aligned}
\end{equation}
($n>2$) where
$${\cal V}_{0}\equiv 0\quad ,\ {\cal V}_{1}\equiv 1\quad ,\  \pa_{i}\equiv \pa_{{\cal V}_i}\quad ,\ \pa^2_{i,j}\equiv \pa_{{\cal V}_i} \pa_{{\cal V}_j}$$
and
$a_i,b_i,c_{i,j},f_{i,j},e_i$ are constants that can depend on $n$.
In particular,
$$a_{n-1}=\frac{2}{{(n-1)}^2}\quad ,\quad b_{2}=2\ \,n\quad ,\quad  e_0=n(n-1)\quad ,\quad
e_{j-2}=\frac{n-j+1}{{(j-1)}^2}\ .$$
  \item $\Delta_{n,q}=\Delta_{n,q}(\{{\cal V}_k\},q_1,q_2,\ldots,q_{w})$, $w=(n-1)(n-2)/2$: for arbitrary $d$, this operator annihilates any volume-like function, namely, $\Delta_{n,q}\,f(\{{\cal V}_k\})=0$\ . We were unable to find explicitly other constant for arbitrary $n$\ .
\end{itemize}
The operator (\ref{DVn}) is $sl(n, {\bf R})$-Lie-algebraic and is gauge-equivalent to a $(n-1)$-dimensional Schr\"odinger operator in a curved space. For this operator ${\De}_{n,g}$, the reduction from $d=n-1$ to $d=n-2$ simply corresponds to the condition ${\cal V}_{n}=0$ while the reduction to $d=n-3$ occurs when ${\cal V}_{n}={\cal V}_{n-1}=0$ and so on. All the limits from $d\geq n-1$ to $d=\tilde{d}<n-1$ are geometrically transparent and, more importantly, ${\De}_{n,g}$ remains algebraic. The form of (\ref{DVn}) implies the existence of a subset of eigenfunctions in the form of a global factor times a polynomial solution in the variables $\{{\cal V}_k\}$. These geometrical variables can be generalized to the case of non equal masses.


\newpage


\begin{thebibliography}{99}

\bibitem{TME3-3}
          A.~Turbiner, W.~Miller,~Jr. and M.~A.~Escobar-Ruiz,\\
         \textit{Three-body problem in $3D$ space: ground state, (quasi)-exact-solvability},\\
         {\it J. Phys. \bf A50} (2017) 215201 (19pp)

\bibitem{TME3-d}
         A.~Turbiner, W.~Miller,~Jr. and M.~A.~Escobar-Ruiz,\\
         \textit{Three-body problem in $d$-dimensional space: ground state, (quasi)-exact-solvability},\\
         {\it Journal of Math. Phys. \bf A59} (2018) 022108 (29pp)

\bibitem{MTE:2018}
        W.~Miller,~Jr., A.V.~Turbiner and M.A.~Escobar-Ruiz,  \\
        \textit{The quantum $n$-body problem in dimension $d\geq n-1$: ground state },\\
        {\it J. Phys. \bf A51} (2018) 205201 (25pp)

\bibitem{RT:1996}
         W.~R{\"u}hl and A.~V.~Turbiner,\\
         \textit{Exact solvability of the Calogero and Sutherland models},\\
         {\it Mod. Phys. Lett.} {\bf A10} (1995), 2213--2222,
         {\tt hep-th/9506105}

\bibitem{Ter}
        K.A.~Ter-Martirosyan, \\
        at {\textit {Lectures on quantum field theory}}, {\it ITEP, Moscow, circa} 1972
        (unpublished)

\bibitem{Gu}
        X.-Y.~Gu, Z.-Q.~Ma, J.-Q.~Sun\\
        {\textit {Quantum four-body system in $D$ dimensions}},\\
        {\it J. Math. Phys.} {\bf 44} 3763 (2003)

\bibitem{Turbiner:1998}
          A.V.~Turbiner, \\
          \textit{Hidden Algebra of Three-Body Integrable Systems},\\
          {\it Modern Phys.Lett. \bf A13}, 1473-1483 (1998)

\bibitem{KKM}E. G. Kalnins, J. M. Kress, and W. Miller, Jr.\\
 \textit{Separation of variables and Superintegrability: The symmetry of solvable systems},\\ {\it Instititute of Physics, UK, (2018), ISBN: 978-0-7503-1314-8, e-book}

\bibitem{Turbiner:1988}
        A.V.~Turbiner, \\
        \textit{Quasi-Exactly-Solvable Problems and the $sl(2,R)$ algebra},\\
        \textit{Comm.Math.Phys. \bf 118} (1988) 467-474

\bibitem{Turbiner:2016}
        A.V.~Turbiner, \\
        \textit{One-dimensional Quasi-Exactly-Solvable Schr\"odinger equations},\\
        \textit{Phys. Repts. \bf 642} (2016) 1-71

\bibitem{Erdi}
         {\'E}rdi, B{\'a}lint and Czirj{\'a}k, Zal{\'a}n, \\
         \textit{Central configurations of four bodies with an axis of symmetry},\\
         {\em Celestial Mechanics and Dynamical Astronomy \bf 125} (2016) 33-70

\bibitem{Hampton}
         M. Hampton and R. Moeckel, \\
         \textit{Finiteness of relative equilibria of the four-body problem},\\
         {\it Inventiones mathematicae \bf 163} (2006) 289-312

\bibitem{Albouy}
         A. Albouy, \\
         \textit{The symmetric central configurations of four equal masses},\\
         In {\em Hamiltonian Dynamics and Celestial Mechanics,\\ Contemp. Math. \bf 198} (1996)  131-135

\bibitem{Calogero:1969}
          F.~Calogero,\\
          \textit{Solution of a three-body problem in one dimension},\\
          {\it J. Math. Phys. \bf 10} (1969),  2191--2196;\\
          \textit{Solution of the one-dimensional $N$-body problem with quadratic
          and/or inversely quadratic pair potentials},\\
          {\it J. Math. Phys.} {\bf 12} (1971) 419--436

\bibitem{Green}
          H.S.~Green,\\ \textit{Structure and energy levels of light nuclei},\\
          {\it Nuclear Physics \bf 54}, 505 (1964)

\bibitem{Moshinsky}
          M.~Moshinsky and C.~Kittel,\\
          \textit{How good is the Born-Oppenheimer approximation?}\\
          {\it Proc. Natl. Acad. Sci. U.S.A. \bf 60}, 1110 (1968)

\bibitem{MMA}
          A.~Minzoni, M.~Rosenbaum and A.~Turbiner,\\
          \textit{Quasi-exactly-solvable many-body problems}\\
          {\it Mod. Phys. Lett. A \bf 11}, 1977-1984 (1996)



\end{thebibliography}
\end{document}